\documentclass{emulateapj}

\shorttitle{Apsidal motion and LC solution for 18 SMC eccentric EBs} \shortauthors{Zasche et al.}

\begin{document}

\title{\textbf{Apsidal motion and light a curve solution \\ for eighteen SMC eccentric eclipsing binaries}\thanks{Based on data collected with the Danish 1.54-m telescope at the ESO La Silla Observatory.
       Tables \ref{minima} are only available in electronic form at the CDS via anonymous ftp to cdsarc.u-strasbg.fr (130.79.128.5) or via
       http://cdsweb.u-strasbg.fr/cgi-bin/qcat?J/A+A/ } }

\author{P. Zasche~\altaffilmark{1}
    \and M. Wolf~\altaffilmark{1}
    \and J. Vra\v{s}til~\altaffilmark{1}
    \and J. Li\v{s}ka~\altaffilmark{2}
    \and M. Skarka~\altaffilmark{2}
    \and M. Zejda~\altaffilmark{2} }

\email{zasche@sirrah.troja.mff.cuni.cz}

\affil{
 \altaffilmark{1} Astronomical Institute, Charles University in Prague, Faculty of Mathematics and Physics, \\
 CZ-180~00 Praha 8, V~Hole\v{s}ovi\v{c}k\'ach 2, Czech Republic\\
 \altaffilmark{2} Department of Theoretical Physics and Astrophysics, Masaryk University, Kotl\'a\v{r}sk\'a 2, 611 37 Brno, Czech Republic }

\date{Received \today}

\begin{abstract}
{Aims: The Danish 1.54-meter telescope at the La Silla observatory was used for photometric
monitoring of selected eccentric eclipsing binaries located in the Small Magellanic Cloud. The new
times of minima were derived for these systems, which are needed for accurate determination of the
apsidal motion. Moreover, many new times of minima were derived from the photometric databases OGLE
and MACHO. Eighteen early-type eccentric-orbit eclipsing binaries 
were studied. }

 {Methods: Their $O-C$\ diagrams of minima timings were analysed and the parameters
of the apsidal motion were obtained. The light curves of these eighteen binaries were analysed
using the program {\sc PHOEBE}, giving the light curve parameters. For several systems the
additional third light also was detected. }

 {Results: We derived for the first time and
significantly improved the relatively short periods of
apsidal motion 
from 19 to 142 years for these systems. 
The relativistic effects are weak, up to 10\, \% of the total apsidal motion rate. For one system
(OGLE-SMC-ECL-0888), the third-body hypothesis was also presented, which agrees with high value of
the third light for this system detected during the light curve solution.}
 {}
 \end{abstract}


\section{Introduction}

Other galaxies have become the most prominent battlefields in current astrophysical research,
mainly due to the large and long-lasting photometric surveys. These surveys like MACHO or OGLE have
discovered thousands of new eclipsing binaries in the Magellanic Clouds, hence, we know only about
twice more eclipsing binaries in our own Milky Way than in other galaxies (see
Pawlak et al. 2013, or Graczyk et al. 2011).                                                

On the other hand, the chemical composition of the Magellanic Clouds differs from that of the solar
neighborhood (e.g. Ribas 2004), and the study of the massive and metal-deficient stars in the SMC
checks our evolutionary models for these abundances. All eclipsing binaries analysed here have
properties that make them important astrophysical laboratories for studying the structure and
evolution of massive stars (Ribas 2004).

Eccentric eclipsing binaries (hereafter EEBs) with an apsidal motion can provide us with an
important observational test of theoretical models of stellar structure and evolution. A long-term
collection of the times of EEBs minima observed for several years throughout the apsidal motion
cycle and a consecutive detailed analysis of the period variations of EEB can be performed,
yielding both the orbital eccentricity and the period of rotation of the apsidal line with high
accuracy (Gim\'enez 1994). Many different sets of stellar evolution models have been published in
recent years, such as for Maeder (1999), or Claret (2005); however, to distinguish between them and
to test, which one is more suitable, it is still rather difficult. The internal structure
constants, as derived from the apsidal motion analysis, could serve as one independent criterion.
On the other hand, only stellar parameters for EEBs with an accuracy of 1\,\% can be used to
discriminate between the models.

Here, we analyse the observational data and rates of apsidal motion for eighteen SMC detached
eclipsing systems. All these systems are early-type objects, having eccentric orbits, which also
exhibits an apsidal motion. Similar studies of LMC EEBs have been presented by Michalska \&
Pigulski (2005), by Michalska (2007), and recently also by Zasche \& Wolf (2013). As far as we
know, only several eclipsing binaries with apsidal motion were analysed in SMC galaxy until now:
SC3 139376, SC5 311566 (Graczyk 2003), and nine other systems by North et al. (2010).

\section{Observations of minimum light}

Monitoring of faint EEBs in external galaxies became almost routine nowadays with quite moderate
telescopes of 1 - 2m class, which are equipped with a modern CCD camera. However, a large amount of
observing time is needed, which is usually unavailable at larger telescopes. During the last two
observational seasons, we have accumulated 2660 photometric observations and derived 29 precise
times of minimum light for selected eccentric systems. New CCD photometry was obtained at the La
Silla Observatory in Chile, where the 1.54-m Danish telescope (hereafter DK154) with the CCD camera
and \textit{R} filter was used (remotely from the Czech Republic).

All CCD measurements were reduced in a standard way using the bias frames and then the flat fields.
The comparison star was chosen to be close to the variable one and with similar spectral type. A
synthetic aperture photometry and astrometry software developed by M.~Velen and P.~Pravec {\sc
Aphot}, was routinely used for reducing the data. No correction for differential extinction was
applied because of the proximity of the comparison stars to the variable and the resulting
negligible differences in air mass and their similar spectral types.

The new times of primary and secondary minima and their respective errors were determined by the
classical Kwee-van Woerden (1956) method or by our new approach (see the section \ref{Method}). All
new times of minima are given in the appendix Tables \ref{minima}.

\section{Photometry and light curve modelling}

The core of our analysis lies on the huge photometric data sets, as obtained during the {\sc Macho}
(Faccioli et al. 2007), {\sc Ogle II} (Wyrzykowski et al. 2004), and {\sc Ogle III} (Graczyk et al.
2011) surveys. These photometric data were used both for minima time analysis and for light curve
analysis. The method of how the individual times of minima for the particular system were computed
is presented in section \ref{Method}. Our new observations obtained at the Danish 1.54-m telescope
were used only for deriving the times of minima for the selected targets.

The analysis of the light curves (hereafter LC) for the systems was carried out using the program
{\sc PHOEBE}, ver. 0.31a (Pr{\v s}a \& Zwitter 2005), which is based on the Wilson-Devinney
algorithm (Wilson \& Devinney 1971) and its later modifications, but some of the parameters have to
be fixed during the fitting process. The albedo coefficients $A_i$ remained fixed at value 1.0, the
gravity darkening coefficients $g_i = 1.0$. The limb darkening coefficients were interpolated from
the van Hamme's tables (van Hamme 1993),  
and the synchronicity parameters ($F_i$) were also kept fixed at values of $F_i = 1$. The
temperature of the primary component was derived from the photometric indices or other sources (see
below). The problematic issue of the mass ratio was solved by fixing $q=1$ because no spectroscopy
for most of these selected systems exists, and for detached eclipsing binaries the LC solution is
almost insensitive to the photometric mass ratio (see e.g. Terrell \& Wilson 2005).

\section{Methods used for the analysis} \label{approach}

\subsection{Apsidal motion analysis}

For the analysis, we used the approach as presented below.

 \begin{enumerate}
   \item At the beginning, all of the available photometric data were analysed, resulting in a set
   of minima times. Preliminary apsidal motion parameters were derived (with the assumption $i=90^\circ$).\\[-1mm]

   \item Secondly, the eccentricity ($e$), argument of periastron ($\omega$), and apsidal motion
   rate ($\dot \omega$) that resulted from the apsidal motion analysis were used for the preliminary
   light curve analysis.\\[-1mm]

   \item As the third step, the inclination ($i$) from the LC analysis was used for the final apsidal motion
   analysis.\\[-1mm]

   \item Finally, the resulted $e$, $\omega$, and $\dot \omega$ values from the apsidal motion
   analysis were used for the final LC analysis.\\[-1mm]
 \end{enumerate}

Moreover, this simple approach was a bit complicated because the minima times were also derived
using the light curve template (see the AFP method in Section \ref{Method}). Hence, the LC solution
from step 2 allows us to derive the better times of minima for the step 3. The whole process run
iteratively until the changes are negligible (usually it was enough to run these four steps two
times).

%

The \oc\ diagrams of all available times of minima were analysed using the method presented by
Gim\'enez \& Garc\'{\i}a-Pelayo (1983). This is a weighted least-squares iterative procedure,
including terms in the eccentricity up to the fifth order. There are five independent variables
$(T_0, P_s, e, \dot{\omega}, \omega_0)$ determined in this procedure. The periastron position
$\omega$ is given by the linear equation

\medskip
\noindent $ \omega = \omega_0 + \dot{\omega}\ E $,

\medskip
\noindent where $\dot{\omega}$ is the rate of periastron advance, $E$ is the epoch, and the
position of periastron for the zero epoch $T_0$ is denoted as $\omega_0$. The relation between the
sidereal and the anomalistic period, $P_s$ and $P_a$, is given by

\medskip
\noindent $ P_s = P_a \,(1 - \dot{\omega}/360^\circ) $

\medskip
\noindent and the period of apsidal motion by $ U = 360^\circ P_a/\dot{\omega} $.

All new precise CCD times of minima were used with a weight of 10 in our computation; some of our
less precise measurements were weighted by a factor of five, while the poorly covered minima were
given a weight of 1.

\subsection{Method of minima fitting} \label{Method}

We developed and routinely used a method for deriving the times of minima for selected stars
observed during the {\sc MACHO} and {\sc OGLE} surveys. This semi-automatic fitting procedure
(hereafter AFP) has harvested the fact that the number of data points obtained during these two
photometric surveys is large (typically thousands of data points) but obtained during many orbital
revolutions of the close pair (a so-called sparse photometry).

Therefore, we can construct the phased light curve of the eclipsing binary in different time
epochs. If the apsidal motion is prominent in the system, the shape of the light curve also
slightly varies between the different epochs.

The first step is to divide the whole data set of photometry into several different ``subsets'',
which are used for constructing the individual light curves. Then, we usually choose the data set
closest to the half of the time interval covered with observations and use these data points for
constructing the light curve to be analysed.

Then, this light curve is analysed using the {\sc PHOEBE} code, and the theoretical light curve
template is being constructed. This LC model is then being used for deriving the individual times
of minima easily by fitting this phased light curve to the phased light curves for the individual
data sets. The best fit is obtained with the simplex algorithm and the least squares fitting method
by only shifting the theoretical and observed light curve in two axis (phase and magnitude). If the
star has constant magnitude over the whole time range of our data, there is no need to fit the
magnitude shift, and only one free parameter is computed. When we find the best fit, then the times
of minima are computed easily according to the ephemerides for a particular data set. Of course,
for eccentric orbit binaries, both primary and secondary minima are being computed separately.

For the input, there are the data points, the time intervals, the ephemerides, and also parameters
of the method. These are the duration of eclipse (how large portion of the phase curve around
minima is being used for computing), minimum number of data points (if lower, the minimum is not
computed), and the depth of minima. If $1/5$ of the depth of minima is covered with data points,
then this particular minimum is being computed.

Hence, by using this technique, we can usually obtain both primary and secondary minima for each
data subset from an original photometry file. Moreover, this method can also be used in these
cases, where the minimum is covered only very poorly, or only a descent to the minimum is covered.
In these cases, the classical Kwee-van Woerden method would not work properly, so we can obtain
more useful data points. On the other hand, we would like to emphasize that the method is suitable
only for systems with low eccentricity, where the shape of the light curve is changing only
slightly. Otherwise, we have to construct a separate light curve template for each of the data
subset.

\begin{figure}
 \includegraphics[width=0.48\textwidth]{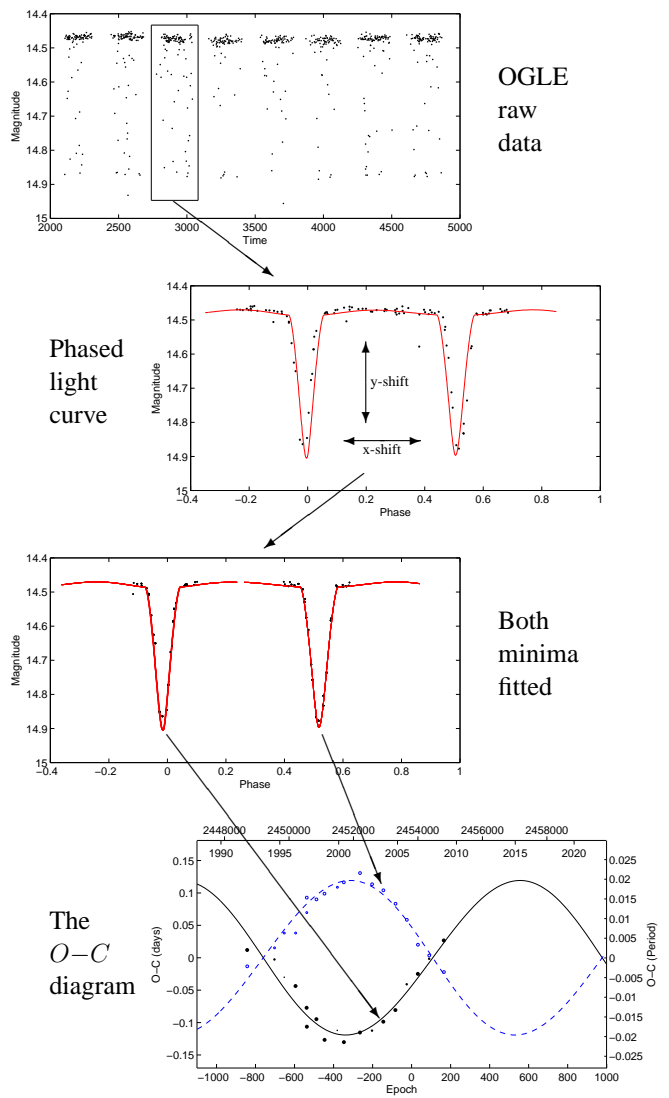}
 \caption[]{How the AFP method works.} \label{MethodObr}
\end{figure}

The whole method is graphically shown in Fig. \ref{MethodObr}, where an illustrative example of
OGLE-SMC-ECL-0720 is being presented. All of the derived times of minima are stored in the
online-only Tables \ref{minima}. There are also given the errors of individual minima times, which
are being computed also by AFP in the following way. The set of different solutions was computed
for a particular minimum with different parameters of the code (length of interval around each
minimum used for the analysis, number of data points according to their precision, etc.), yielding
a set of times of minima, which is usually more than 10. From these minima data set, an average and
its variance were computed,. The variance is then taken as an approximate error estimation for the
particular minimum.

\begin{table*}[h!]
\caption{Relevant information for the analysed systems.}  \label{InfoSystems}  \scriptsize
\begin{tabular}{lcccccccccl}
   \hline\hline\noalign{\smallskip}
 System   & OGLE II$^{*}$ &  MACHO     &          RA           &             DE                              & $I_{\rm max}$ & $(B-V)$ & $(U-B)$ & $(B-V)_0$ & Sp.type$^{**}$ & Ref. \\
  \hline\noalign{\smallskip}
 $\#$0720 & SC3 139376 & 213.15620.12  & 00$^h$44$^m$08$^s$.67 & -73$^\circ$14$^\prime$18$^{\prime\prime}$.5 &  14$^{\rm m}\!\!.$47       &  -0.16  & -0.91   & -0.26     & B0(IV)       & 1 , 2      \\
 $\#$0781 & SC3 157218 & 212.15624.89  & 00$^h$44$^m$39$^s$.73 & -72$^\circ$59$^\prime$58$^{\prime\prime}$.5 &  17$^{\rm m}\!\!.$03       &  -0.10  & -0.78   & -0.23     & \emph{B2}    & 1          \\
 $\#$0888 & SC4 29231  & 212.15680.18  & 00$^h$45$^m$30$^s$.72 & -73$^\circ$03$^\prime$29$^{\prime\prime}$.7 &  15$^{\rm m}\!\!.$52       &  -0.09  & -0.82   & -0.25     & O9V          & 1 , 2      \\
 $\#$1001 &            & 208.15744.1836& 00$^h$46$^m$11$^s$.29 & -72$^\circ$35$^\prime$17$^{\prime\prime}$.3 &  18$^{\rm m}\!\!.$44       &  -0.31  & -0.58   & -0.12     & \emph{late B/early A} & 3 \\
 $\#$1298 & SC4 163754 & 212.15848.1258& 00$^h$47$^m$52$^s$.73 & -73$^\circ$16$^\prime$34$^{\prime\prime}$.0 &  17$^{\rm m}\!\!.$22       &  -0.04  & -0.89   & -0.29     & \emph{B0}    & 1          \\
 $\#$1407 &            & 208.15861.734 & 00$^h$48$^m$19$^s$.26 & -72$^\circ$21$^\prime$40$^{\prime\prime}$.2 &  17$^{\rm m}\!\!.$02       &  -0.04  & -0.67   & -0.21     & \emph{B3}    & 1          \\
 $\#$2186 & SC5 311566 & 208.16083.86  & 00$^h$51$^m$34$^s$.84 & -72$^\circ$45$^\prime$46$^{\prime\prime}$.4 &  16$^{\rm m}\!\!.$06       &  -0.04  & -0.79   & -0.25     & B0+B0-3      & 1 , 4      \\
 $\#$2225 & SC6 72782  & 208.16084.117 & 00$^h$51$^m$41$^s$.80 & -72$^\circ$41$^\prime$06$^{\prime\prime}$.1 &  16$^{\rm m}\!\!.$71       &  -0.128 &         & -0.22     & \emph{B3}    & 5 , 6      \\
 $\#$2251 & SC6 61418  &               & 00$^h$51$^m$46$^s$.64 & -72$^\circ$51$^\prime$21$^{\prime\prime}$.7 &  16$^{\rm m}\!\!.$23       &  -0.14  & -0.91   & -0.27     & \emph{B1}    & 1          \\
 $\#$2524 & SC6 158178 & 208.16141.60  & 00$^h$52$^m$42$^s$.32 & -72$^\circ$41$^\prime$27$^{\prime\prime}$.9 &  16$^{\rm m}\!\!.$53       &  -0.176 & -0.86   & -0.24     & \emph{B2}    & 3          \\
 $\#$2534 &            & 208.16147.22  & 00$^h$52$^m$43$^s$.85 & -72$^\circ$18$^\prime$08$^{\prime\prime}$.6 &  16$^{\rm m}\!\!.$57       &  -0.04  & -0.81   & -0.26     & \emph{B1}    & 1          \\ 
 $\#$3594 & SC7 255621 & 207.16428.1423& 00$^h$57$^m$26$^s$.41 & -72$^\circ$36$^\prime$46$^{\prime\prime}$.2 &  16$^{\rm m}\!\!.$26       &  -0.21  & -0.72   & -0.19     & B1+B1-3      & 1 , 4      \\
 $\#$3677 & SC8 52815  & 207.16490.6   & 00$^h$57$^m$49$^s$.25 & -72$^\circ$16$^\prime$55$^{\prime\prime}$.7 &  15$^{\rm m}\!\!.$11       &  -0.16  & -0.84   & -0.24     & \emph{B2}    & 1 , 7      \\ 
 $\#$3951 & SC8 160725 &               & 00$^h$59$^m$14$^s$.98 & -72$^\circ$11$^\prime$35$^{\prime\prime}$.3 &  15$^{\rm m}\!\!.$90       &  -0.18  & -0.84   & -0.24     & B1V          & 8 , 9      \\
 $\#$4955 & SC10 94636 & 206.16886.52  & 01$^h$04$^m$59$^s$.18 & -72$^\circ$25$^\prime$29$^{\prime\prime}$.3 &  17$^{\rm m}\!\!.$11       &  -0.16  & -0.68   & -0.19     & \emph{B3}    & 1          \\ 
 $\#$5233 &            & 206.17061.14  & 01$^h$07$^m$12$^s$.54 & -72$^\circ$11$^\prime$42$^{\prime\prime}$.0 &  15$^{\rm m}\!\!.$34       &  -0.08  & -0.91   & -0.28     & \emph{B0}    & 1          \\
 $\#$5422 & SC11 111907& 206.17170.8   & 01$^h$08$^m$45$^s$.74 & -72$^\circ$31$^\prime$22$^{\prime\prime}$.4 &  14$^{\rm m}\!\!.$99       &  -0.22  & -0.93   & -0.26     & \emph{B1}    & 1          \\ 
 $\#$5434 & SC11 118966& 206.17173.10  & 01$^h$08$^m$50$^s$.47 & -72$^\circ$17$^\prime$26$^{\prime\prime}$.1 &  15$^{\rm m}\!\!.$56       &  -0.11  & -0.86   & -0.26     & \emph{B1}    & 1          \\ 
 \noalign{\smallskip}\hline
\end{tabular}
\scriptsize Note: [*] - The full name from OGLE II survey should be OGLE SMC-SCn nnnnnn, [**] -
Spectral types given in italics were only estimated from the photometric indices for the first time
in the present paper. References: [1] - Massey (2002), [2] - Evans et al. (2004), [3] - Zaritsky et
al. (2002), [4] - Hilditch et al. (2005), [5] - Udalski et al. (1998), [6] - Massey et al. (1995),
[7] - Bonanos et al. (2010), [8] - Massey et al. (1989), [9] - Massey et al. (2012).
\end{table*}

\begin{table*}[h!]
\caption{Light curve parameters for the analysed systems.} \label{LCparam} \tiny
\begin{tabular}{lcccccccccc}
  \hline\hline\noalign{\smallskip}
   System          &   $T_1$ [K]   &  $T_2$ [K]  &  $i$ [deg]   &  $\Omega_1$   &  $\Omega_2$   &  $L_1$ [\%]  &  $L_2$  [\%] & $L_3$ [\%] \\
  \hline\noalign{\smallskip}
 $\#$0720 & 31500 (fixed) & 31700 (400) & 84.57 (0.30) & 5.524 (0.061) & 7.356 (0.085) & 66.87 (0.83) & 33.13 (0.68) &  0 \\
 $\#$0781 & 23100 (fixed) & 16700 (300) & 85.08 (0.18) & 7.265 (0.086) & 8.690 (0.120) & 71.90 (1.25) & 28.10 (0.98) &  0 \\
 $\#$0888 & 33200 (fixed) & 38100 (1100)& 77.77 (0.38) & 6.119 (0.101) & 6.128 (0.112) & 22.01 (1.85) & 26.94 (3.46) & 51.05 (4.98) \\
 $\#$1001 & 11000 (fixed) &  9700 (500) & 79.13 (0.92) & 5.064 (0.204) & 6.661 (0.397) & 67.48 (1.37) & 32.52 (1.02) &  0 \\
 $\#$1298 & 30000 (fixed) & 19900 (700) & 74.68 (0.45) & 6.364 (0.138) & 6.701 (0.174) & 70.46 (1.32) & 29.54 (0.77) &  0 \\
 $\#$1407 & 19000 (fixed) & 19100 (700) & 75.60 (0.50) & 5.907 (0.147) & 6.833 (0.177) & 56.26 (1.06) & 39.10 (1.00) &  4.64 (1.87) \\
 $\#$2186 & 30100 (fixed) & 28500 (300) & 87.02 (0.22) & 6.843 (0.058) & 8.040 (0.097) & 62.73 (3.14) & 36.00 (1.25) &  4.03 (2.59) \\
 $\#$2225 & 11600 (fixed) &  7100 (200) & 80.33 (0.53) & 5.678 (0.100) &11.334 (0.490) & 60.98 (2.87) &  4.58 (0.57) & 34.45 (1.55) \\
 $\#$2251 & 26200 (fixed) & 30600 (900) & 79.62 (0.26) & 6.149 (0.080) &10.184 (0.184) & 48.35 (3.02) & 17.91 (1.26) & 33.73 (4.05) \\
 $\#$2524 & 23100 (fixed) & 24600 (600) & 83.32 (0.37) & 6.128 (0.085) & 7.165 (0.112) & 56.63 (1.26) & 41.53 (2.03) &  1.85 (5.76) \\
 $\#$2534 & 26200 (fixed) & 18300 (300) & 76.23 (0.28) & 5.690 (0.050) & 6.150 (0.052) & 61.67 (1.79) & 28.64 (1.58) &  9.69 (2.63) \\
 $\#$3594 & 25500 (fixed) & 22000 (200) & 83.60 (0.41) & 7.095 (0.040) & 6.668 (0.055) & 50.50 (1.56) & 48.00 (1.03) &  1.49 (1.96) \\
 $\#$3677 & 23100 (fixed) & 25100 (300) & 73.60 (0.22) & 5.029 (0.029) & 7.833 (0.064) & 74.51 (1.07) & 25.49 (0.94) &  0 \\
 $\#$3951 & 26200 (fixed) & 24400 (200) & 78.50 (0.24) & 6.699 (0.055) & 6.789 (0.056) & 53.76 (0.97) & 46.24 (1.16) &  0 \\
 $\#$4955 & 19000 (fixed) & 17400 (500) & 80.01 (0.31) & 7.868 (0.165) & 8.591 (0.180) & 58.97 (0.75) & 41.03 (0.79) &  0 \\
 $\#$5233 & 30000 (fixed) & 29400 (300) & 80.52 (0.21) & 8.762 (0.098) & 7.843 (0.093) & 44.09 (1.02) & 55.91 (0.93) &  0 \\
 $\#$5422 & 26200 (fixed) & 20400 (300) & 79.18 (0.33) & 7.076 (0.098) & 7.070 (0.102) & 54.69 (2.34) & 36.51 (4.57) &  8.80 (7.82) \\
 $\#$5434 & 26200 (fixed) & 24200 (400) & 71.72 (0.40) & 5.400 (0.061) & 5.516 (0.059) & 53.49 (1.14) & 43.86 (1.62) &  2.65 (2.01) \\
 \noalign{\smallskip}\hline
\end{tabular}
\end{table*}

\begin{table*}[h!]
\caption{The parameters of the apsidal motion for the individual systems.} \label{OCparam}
\scriptsize
\begin{tabular}{lcccccccc}
\hline\hline\noalign{\smallskip}
  System           & $T_0 - 2400000$ [HJD] &  $P_s$ [days]   &   $e$       & $\dot{\omega}$ [deg$/\rm{cycle}$] & $\omega_0$ [deg] & $U$ [yr] \\
 \noalign{\smallskip}\hline\noalign{\smallskip}
 $\#$0720 & 53803.390 (21)        & 6.052322 (48) & 0.062 (16)& 0.2070 (300)                      & 67.6 (5.0)       & 28.8 (5.5)  \\ 
 $\#$0781 & 52745.417 (32)        & 3.299923 (48) & 0.310 (75)& 0.0301 (37)                       & 244.4 (10.3)     & 108.1 (15.1)\\ 
 $\#$0888 & 53470.954 (15)        & 1.918337 (11) & 0.143 (34)& 0.0474 (193)                      & 89.3 (4.7)       & 39.9 (11.6) \\ 
 $\#$1001 & 53090.7643 (35)       & 1.1621122 (18)& 0.072 (14)& 0.0601 (69)                       & 94.5  (7.8)      & 19.0 (2.4)  \\ 
 $\#$1298 & 53501.194 (14)        & 1.7532121 (99)& 0.219 (42)& 0.0677 (93)                       & 130.0 (4.2)      & 25.5 (4.1)  \\ 
 $\#$1407 & 53470.497 (13)        & 2.100755  (11)& 0.151 (47)& 0.0427 (120)                      & 115.9 (3.9)      & 46.3 (18.0) \\ 
 $\#$2186 & 53470.314 (15)        & 3.291316  (20)& 0.227 (82)& 0.0258 (42)                       & 91.9 (2.6)       & 125.7 (24.5)\\ 
 $\#$2225 & 53089.914 (10)        & 1.491721  (8) & 0.187 (48)& 0.0351 (121)                      & 291.9 (8.8)      & 41.9 (21.8) \\ 
 $\#$2251 & 54179.695 (25)        & 2.336038 (33) & 0.271 (22)& 0.0484 (132)                      & 99.8 (5.6)       & 47.6 (17.8) \\ 
 $\#$2524 & 53471.664 (24)        & 2.169236 (23) & 0.263 (68)& 0.0475 (92)                       & 81.5 (6.5)       & 45.0 (10.8) \\ 
 $\#$2534 & 53277.1246 (72)       & 2.2967384 (72)& 0.078 (24)& 0.0357 (57)                       & 265.0 (3.4)      & 63.3 (11.9) \\ 
 $\#$3594 & 53280.315 (41)        & 4.330333 (80) & 0.194 (69)& 0.0300 (63)                       & 238.1 (9.9)      & 142.1 (38.9)\\ 
 $\#$3677 & 53278.036 (52)        & 5.241539 (117)& 0.153 (55)& 0.0554 (133)                      & 40.2 (4.6)       & 93.3 (33.6) \\ 
 $\#$3951 & 53277.2672 (75)       & 3.104291 (17) & 0.092 (20)& 0.0476 (165)                      & 92.9 (4.0)       & 64.3 (33.9) \\ 
 $\#$4955 & 54022.771 (52)        & 2.772183 (53) & 0.338 (48)& 0.0239 (84)                       & 143.3 (5.0)      & 114.4 (51.5)\\ 
 $\#$5233 & 52746.459 (45)        & 5.068362 (103)& 0.199 (57)& 0.1915 (321)                      & 7.0 (8.7)        & 26.1 (5.2)  \\ 
 $\#$5422 & 53656.966 (26)        & 3.040295 (31) & 0.199 (56)& 0.0301 (83)                       & 318.1 (6.0)      & 99.4 (37.9) \\ 
 $\#$5434 & 53478.7191 (72)       & 2.886936 (9)  & 0.051 (16)& 0.0747 (214)                      & 129.6 (3.4)      & 38.1 (15.2) \\ 
 \noalign{\smallskip}\hline
\end{tabular}
\end{table*}

\section{Notes on individual systems}

\begin{figure*}
 \includegraphics[width=\textwidth]{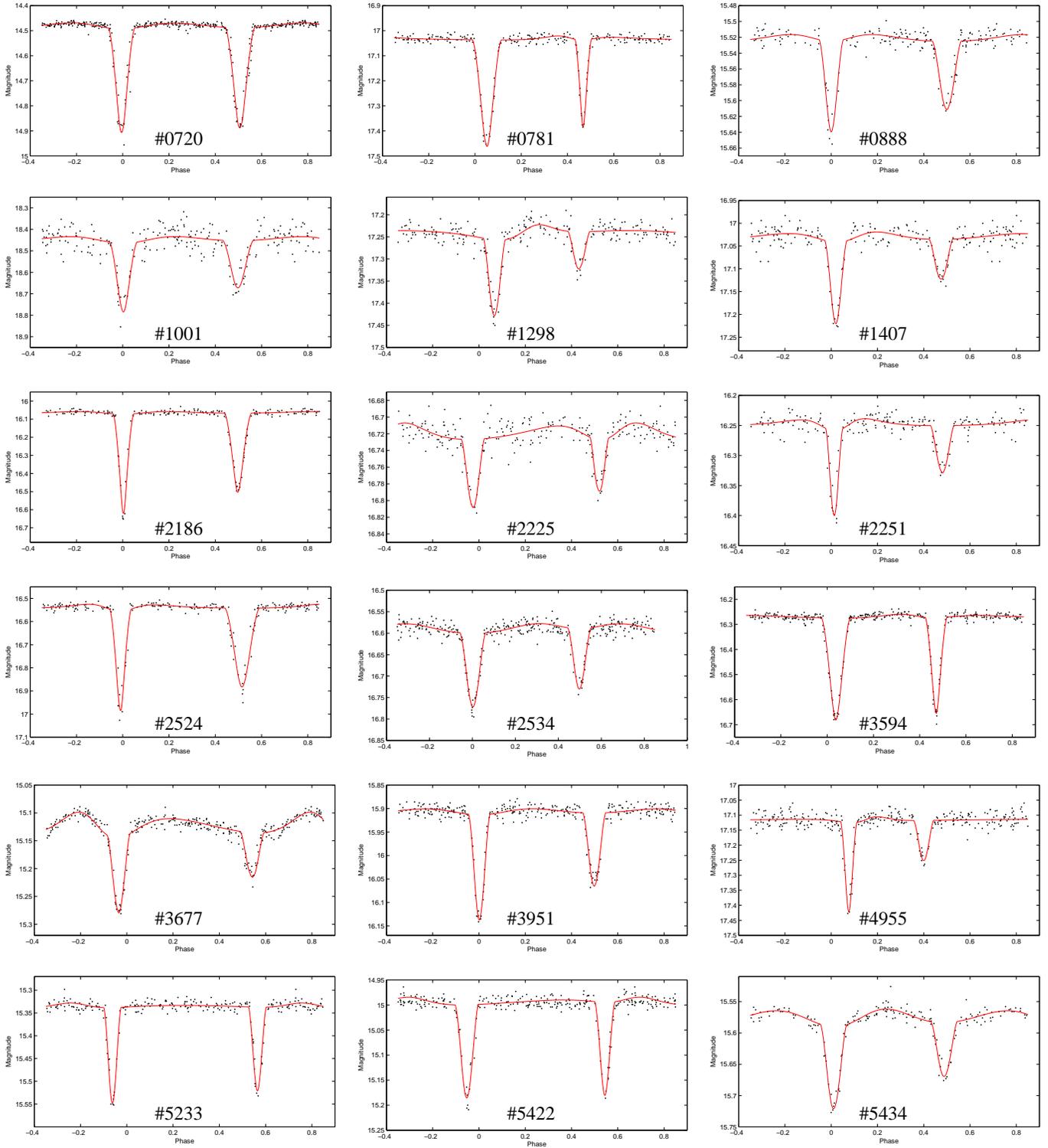}
 \caption[]{Light curves of the analysed systems, the data taken from the OGLE III survey, and the $I$
 filter.} \label{LCs}
\end{figure*}

\begin{figure*}
 \includegraphics[width=\textwidth]{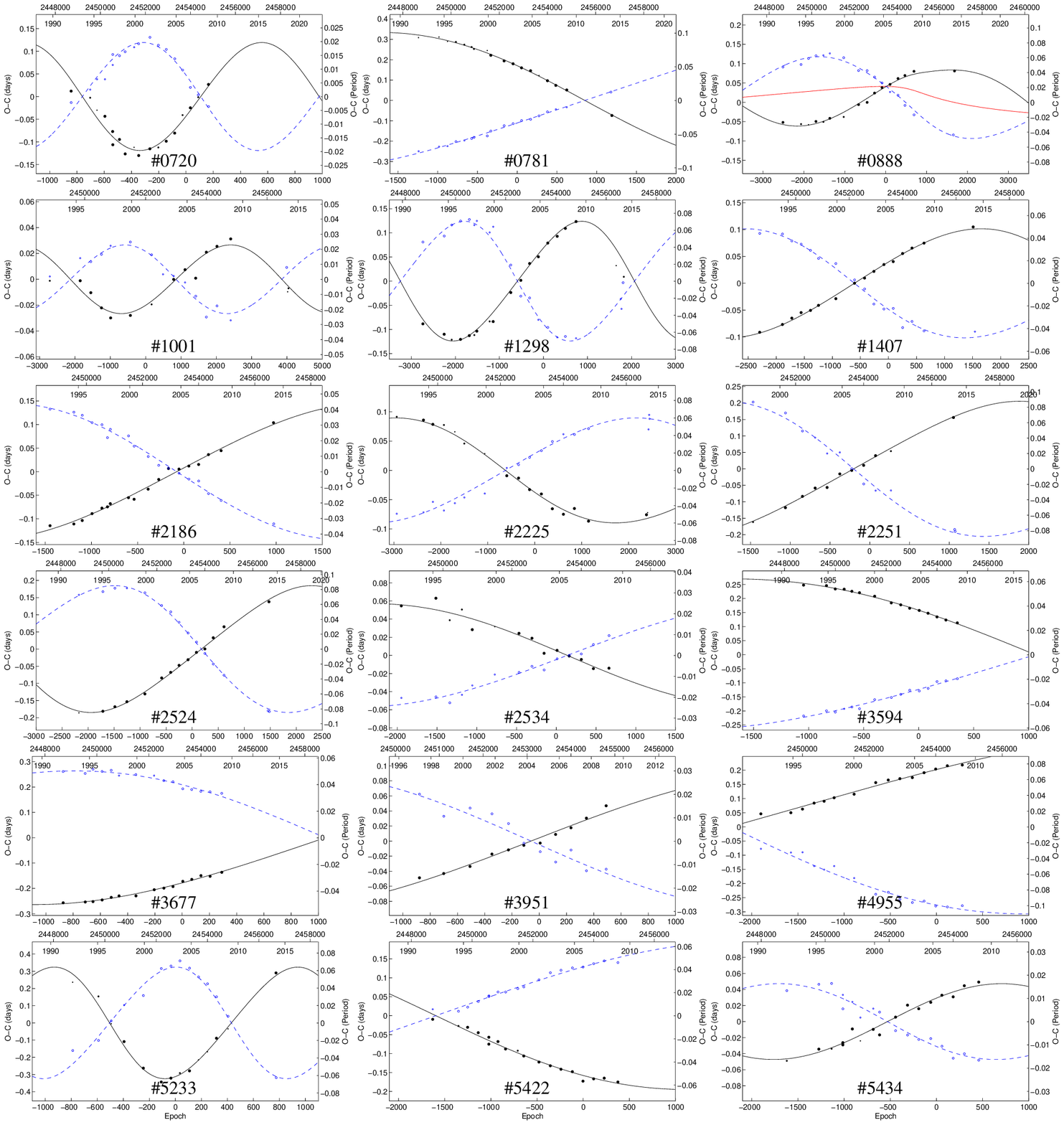}
 \caption[]{\oc\ diagram for the times of minima for the analysed systems. The continuous and dashed
curves represent predictions for the primary and secondary eclipses, respectively. The individual
primary and secondary minima are denoted by dots and open circles, respectively. Larger symbols
correspond to the measurements, which were given higher weights.} \label{OCs}
\end{figure*}

All of the eclipsing systems were analysed using a similar approach, hence we cannot focus on every
star in detail. See Table \ref{InfoSystems} for information and cross-identification of these
stars. The abbreviations of the star names were used for all of the systems for a better brevity.
That is, OGLE-SMC-ECL-0720 was shortened as $\#$0720, etc. Only the most important results are
summarized below. The final light curve fits, and the $O-C$ diagrams are presented in Figs.
\ref{LCs} and \ref{OCs}; the parameters are given in Tables \ref{LCparam} and \ref{OCparam}. The
whole set of eighteen analysed systems can be divided into a few subsets according to available
spectral information.





The largest group in our sample of stars comprise these stars, which were never observed
spectroscopically, hence no spectral classification or radial velocity study was published so far.
These systems are $\#$0781, $\#$1001, $\#$1298, $\#$1407, $\#$2225, $\#$2251, $\#$2524, and
$\#$5233. Most of them were discovered as eclipsing binaries by Udalski et al. (1998), Wyrzykowski
et al. (2004), or Faccioli et al. (2007). Several of them were mentioned as eccentric ones with
apsidal motion in some of the above mentioned papers. Owing to having no information about their
spectra, we only roughly estimated the spectral types from the measurements in photometric filters,
as seen in Table \ref{InfoSystems}. These observations were usually taken from Massey (2002) and
from the dereddened photometric indices the spectral types were estimated (Popper 1980, Ducati et
al. 2001, or Cox 2000). For some of the systems, there resulted a non-negligible third light
contribution (e.g. $\#$2225, $\#$2251).




For some of the systems, the spectral types were published, so we can use them for a better primary
temperature estimation for a subsequent light curve analysis. These systems are $\#$0720, $\#$2534,
$\#$3677, $\#$4955, $\#$5422, and $\#$5434 (to this group of stars, two systems $\#$0888 and
$\#$3951 also belong, but these were given a special focus in the following subsections). These
binaries were also discovered by Udalski et al. (1998) and Faccioli et al. (2007); for some of
them, a short note about their apsidal motion was published. The spectral types for these systems
given by Evans et al. (2004) and Bonanos et al. (2010) are in good agreement with our spectral
types that are estimated from the dereddened photometric indices.




\subsection{OGLE-SMC-ECL-2186}

Two systems, $\#$2186 and $\#$3594, were even published with their light and radial velocity curves
solutions. The first one ($\#$2186) was analysed by Wyithe \& Wilson (2001), who presented a
preliminary light curve solution with an eccentric orbit with $e=0.068$. Graczyk (2003) analysed
the LC of $\#$2186, yielding an eccentricity of 0.251, no third light, and the luminosity ratio of
$L_2/L_1=0.843$. Wyrzykowski et al. (2004) presented a note about its apsidal motion but with no
estimation of its period. Concerning the spectral type, Graczyk (2003) estimated the types of about
O9V+O9V but dealt only with the photometry. Later, Hilditch et al. (2005) published its spectral
type to be B0+B0-3 based on 15 spectra of the star. They also analysed the light curve, yielding a
value of the eccentricity of the orbit to be 0.063. However, their LC solution is not very
convincing due to poor fit of the secondary minimum.

\subsection{OGLE-SMC-ECL-3594}

The second system ($\#$3594) was also studied by Wyithe \& Wilson (2001), who included this star
into their sample of SMC eclipsing binaries with the light curve solution, which result in orbital
inclination of 88.9$^\circ$ and an eccentricity of 0.144. Hilditch et al. (2005) analysed the
system in more detail, resulting in an orbital eccentricity of 0.19 (based on photometry and
spectroscopy together) and the spectral types of both components as B1+B1-3.




\subsection{OGLE-SMC-ECL-0888}

\begin{figure}
\includegraphics[width=0.48\textwidth]{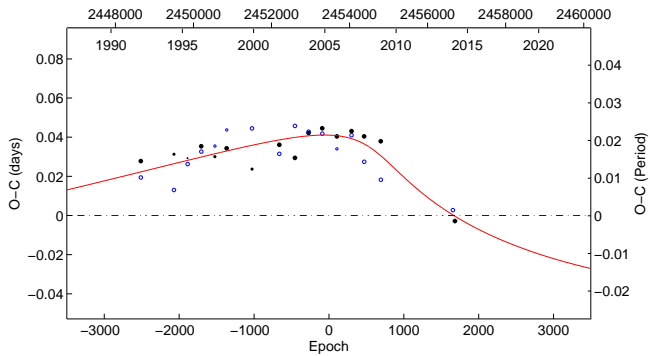}
\caption[ ]{\oc\ diagram of $\#$0888 after subtraction of the apsidal motion term.}
\label{OC0888LITE}
\end{figure}

The object $\#$0888 was first mentioned by Wyrzykowski et al. (2004), who also noted about its
apsidal motion. Its spectral type was derived to be about O9V by Evans et al. (2004). We found that
the pure apsidal motion is not able to describe the $O-C$ diagram in detail, hence another effect
has also to be included. We also tried to fit the parabolic fit to the ephemerides, with the
apsidal motion hypothesis (can be interpreted as a mass transfer between the components, despite
improbable for detached binary). However, this fit was also not very satisfactory. Therefore, we
used a different code that computes the apsidal motion parameters with the third-body orbit (a
so-called 'light travel time' effect), as seen in for example Irwin (1959) or Mayer (1990). Ten
parameters were fitted (five from the apsidal motion, five from the third body hypothesis); thus,
this approach led to an acceptable solution with the lowest sum of squares residuals. The final
parameters of the fit are given in Tables~\ref{OCparam} and \ref{LITE0888}; the complete \oc\
diagrams are shown in Figs.~\ref{OCs} and \ref{OC0888LITE}.

\begin{table}[b]
\caption{Third body orbit parameters for $\#$0888.} \label{LITE0888}
\begin{flushleft}
\begin{tabular}{lcccccc}
\hline\hline\noalign{\smallskip}
Parameter [Unit]     &   Value  \\
\noalign{\smallskip}\hline\noalign{\smallskip}
 $p_3$ [yr]        & 72.1 $\pm$ 28.0 \\
 $A_3$ [day]       & 0.030 $\pm$ 0.011  \\
 $T_3$ [HJD]       & 2454900 $\pm$ 8700  \\
 $e_3$             & 0.709 $\pm$ 0.247  \\
 $\omega_3$ [deg]  & 154.0 $\pm$ 15.6  \\
 \noalign{\smallskip}\hline
\end{tabular}
\end{flushleft}
\end{table}

From the third-body parameters, we could also compute the mass function of the distant component,
which resulted in $f(m_3) = 0.059 \pm 0.015~\mathrm{M_\odot}$. From this value, one can calculate a predicted
minimal mass of the third body (i.e. assuming coplanar orbits $i_3=90^\circ$), which resulted in
$m_{3,min} = 4.9~\mathrm{M_\odot}$. If we propose such a body in the system, one can ask whether it is
detectable somehow in the already obtained data. The period is long for continuous monitoring of
the radial velocity changes, but detecting the third light in the light curve solution would be
promising. Assuming a normal main sequence star, its luminosity would be of about only $L_{3,min} =
1-2$\,\% of the total system luminosity. Such a weak third light would be hardly detectable in our
poor-quality photometric data, but it is worth of try. Hence, we performed a new light curve
solution with a special focus on the value of the third light for a LC solution. The value was
really obtained, and its value is not negligible at all. As one can see from the parameters
presented in Table \ref{LCparam}, the third light represents about one half of the total light.
This finding naturally explains why both the eclipses are so shallow. On the other hand, one can
ask to which body the estimated spectral type of O9V belongs. If the third body is the dominant
source, this is probably the O9V component, but the primary temperature of 33200~K was assumed
using the O9V primary, which now seems to be incorrect. However, having no other relevant
information about the individual spectral types, one cannot easily assume a different primary
temperature. Thus, we can conclude that the third body is probably present and orbits around the EB
pair on orbit which is mildly inclined from the originally assumed 90$^\circ$. It is hard to say
anything more about such a body because of the high errors of the parameters (period, third light,
etc.). More precise photometry or radial velocities would be very welcome for a final confirmation
of our hypothesis.


\subsection{OGLE-SMC-ECL-3951}

\begin{figure}
\includegraphics[width=0.48\textwidth]{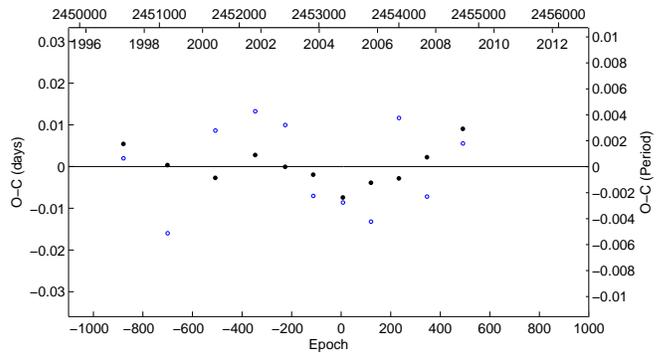}
\caption[ ]{\oc\ diagram of $\#$3951 after subtraction of the apsidal motion term.}
\label{OC3951resid}
\end{figure}

The object $\#$3951 is a part of the SMC open cluster NGC 346. Its eclipsing nature and orbital
period was first presented by Udalski et al. (1998). Later, the star was classified as B1V by
Massey et al. (2012). From the period analysis, a weak quasi-periodic signal also on the residuals
after subtraction of the apsidal motion hypothesis (see Fig. \ref{OC3951resid}) resulted. However,
the variation is still too spurious for any final confirmation yet and we have not even try to fit
the data with any additional variation, as in the previous case.

\section{Discussion and conclusions}

Our study provides the parameters of the apsidal motion for eighteen early-type binary systems
located in the SMC. For most of the binaries, this is the first attempt to estimate the apsidal
motion rates, and the light curve solution. In our own Galaxy there are a few hundreds of apsidal
motion eclipsing binaries known; however, in other galaxies their number is still very limited.
Hence this study still presents an important contribution to the topic. However, for only three
systems from our sample ($\#$1001, $\#$1298, and $\#$5233), the apsidal motion was derived from
adequately large data set covering almost one apsidal period. The relativistic effects for the
selected systems are weak, being up to 10\,\% of the total apsidal motion rate. For the system
$\#$0888, the third body hypothesis was also presented and discussed.

The apsidal motion in EEBs has been used for decades to test evolutionary stellar models. Thus, one
can ask whether our results can be used for deriving the internal structure constants for these
stars in SMC. However, dealing with no radial velocities for most of the systems and with rather
poor data coverage during the apsidal motion period, the parameters are too uncertain and affected
by large errors. For these systems where the apsidal period is well-covered, a detailed
spectroscopic analysis is missing, and vice versa, for systems where the radial velocity study was
performed, the apsidal period has yet to be only poorly covered with data. However, for any testing
of the stellar structure models or for the general relativity tests, the quality of the input data
has to be an order of magnitude better (which implies long-term collection of the observations and
data that covers the whole apsidal period in the following decades). Some of the systems are bright
enough for a spectral monitoring, hence we encourage the observers to obtain new, high-dispersion,
and high-S/N spectroscopic observations. With such data, methods, like spectral disentangling, can
help us construct the radial velocity curves of both components, confirm the apsidal motion
hypothesis, test the stellar structure models, or detect the third bodies, as indicated from our
analysis.

\medskip

\begin{acknowledgements}
We do thank the {\sc MACHO} and {\sc OGLE} teams for making all of the observations easily public
available. This work was supported by the Czech Science Foundation grant no. P209/10/0715, by the
grant UNCE 12 of the Charles University in Prague, and by the grant LG12001 of the Ministry of
Education of the Czech Republic. We are also grateful to the ESO team at the La Silla Observatory
for their help in maintaining and operating the Danish telescope. The following internet-based
resources were used in research for this paper: the SIMBAD database and the VizieR service operated
at the CDS, Strasbourg, France, and the NASA's Astrophysics Data System Bibliographic Services.
\end{acknowledgements}



\begin{appendix} 

\section{Tables of minima}

\begin{table}
 \centering
  \begin{minipage}{95mm}
 \fontsize{1.8mm}{2.4mm}\selectfont
 \caption{List of the minima timings used for the analysis.} \label{minima}
\begin{tabular}{ccclcl}
\hline\hline\noalign{\smallskip}
 Star       &    JD Hel.- &  Error & Type   &  Filter  & Source /     \\
            &   2400000   &  [day] &        &          & Observatory  \\
\noalign{\smallskip}\hline
\noalign{\smallskip}
 OGLE-SMC-ECL-0720 & 48701.29443 & 0.00741 & Prim & B+R & MACHO \\
 OGLE-SMC-ECL-0720 & 48704.29533 & 0.00922 & Sec  & B+R & MACHO \\
 OGLE-SMC-ECL-0720 & 49548.60490 & 0.01197 & Prim & B+R & MACHO \\
 OGLE-SMC-ECL-0720 & 49551.64897 & 0.01236 & Sec  & B+R & MACHO \\
 OGLE-SMC-ECL-0720 & 49851.19352 & 0.01915 & Prim & B+R & MACHO \\
 OGLE-SMC-ECL-0720 & 49854.28789 & 0.02131 & Sec  & B+R & MACHO \\
 OGLE-SMC-ECL-0720 & 50202.21473 & 0.00924 & Prim & B+R & MACHO \\
 OGLE-SMC-ECL-0720 & 50205.32260 & 0.02286 & Sec  & B+R & MACHO \\
 OGLE-SMC-ECL-0720 & 50547.16343 & 0.00657 & Prim & B+R & MACHO \\
 OGLE-SMC-ECL-0720 & 50550.33647 & 0.01600 & Sec  & B+R & MACHO \\
 OGLE-SMC-ECL-0720 & 50849.76208 & 0.00929 & Prim & B+R & MACHO \\
 OGLE-SMC-ECL-0720 & 50852.97305 & 0.00908 & Sec  & B+R & MACHO \\
 OGLE-SMC-ECL-0720 & 51497.34320 & 0.01915 & Prim & B+R & MACHO \\
 OGLE-SMC-ECL-0720 & 51500.59013 & 0.01798 & Sec  & B+R & MACHO \\
 OGLE-SMC-ECL-0720 & 50550.35962 & 0.00592 & Sec  & I   & OGLE II \\
 OGLE-SMC-ECL-0720 & 50553.18661 & 0.00439 & Prim & I   & OGLE II \\
 OGLE-SMC-ECL-0720 & 51101.12700 & 0.00486 & Sec  & I   & OGLE II \\
 OGLE-SMC-ECL-0720 & 51103.92761 & 0.00022 & Prim & I   & OGLE II \\
 OGLE-SMC-ECL-0720 & 51700.32442 & 0.00002 & Sec  & I   & OGLE II \\
 OGLE-SMC-ECL-0720 & 51703.10399 & 0.00089 & Prim & I   & OGLE II \\
 OGLE-SMC-ECL-0720 & 52199.40921 & 0.00391 & Prim & I   & OGLE III \\
 OGLE-SMC-ECL-0720 & 52202.68144 & 0.00649 & Sec  & I   & OGLE III \\
 OGLE-SMC-ECL-0720 & 52574.65616 & 0.01128 & Prim & I   & OGLE III \\
 OGLE-SMC-ECL-0720 & 52577.90814 & 0.00526 & Sec  & I   & OGLE III \\
 OGLE-SMC-ECL-0720 & 52925.70455 & 0.00698 & Prim & I   & OGLE III \\
 OGLE-SMC-ECL-0720 & 52928.93350 & 0.00340 & Sec  & I   & OGLE III \\
 OGLE-SMC-ECL-0720 & 53300.96643 & 0.00535 & Prim & I   & OGLE III \\
 OGLE-SMC-ECL-0720 & 53304.15675 & 0.00206 & Sec  & I   & OGLE III \\
 OGLE-SMC-ECL-0720 & 53652.04092 & 0.01013 & Prim & I   & OGLE III \\
 OGLE-SMC-ECL-0720 & 53655.16657 & 0.00666 & Sec  & I   & OGLE III \\
 OGLE-SMC-ECL-0720 & 53997.03938 & 0.00490 & Prim & I   & OGLE III \\
 OGLE-SMC-ECL-0720 & 54000.11047 & 0.00739 & Sec  & I   & OGLE III \\
 OGLE-SMC-ECL-0720 & 54348.09727 & 0.01183 & Prim & I   & OGLE III \\
 OGLE-SMC-ECL-0720 & 54351.12853 & 0.00173 & Sec  & I   & OGLE III \\
 OGLE-SMC-ECL-0720 & 54802.04975 & 0.00811 & Prim & I   & OGLE III \\
 OGLE-SMC-ECL-0720 & 54805.02717 & 0.00004 & Sec  & I   & OGLE III \\
   \hline
 OGLE-SMC-ECL-0781 & 48650.52017 & 0.00589 & Prim & B+R & MACHO \\
 OGLE-SMC-ECL-0781 & 48651.61198 & 0.00601 & Sec  & B+R & MACHO \\
 OGLE-SMC-ECL-0781 & 49498.60339 & 0.00758 & Prim & B+R & MACHO \\
 OGLE-SMC-ECL-0781 & 49499.71447 & 0.00766 & Sec  & B+R & MACHO \\
 OGLE-SMC-ECL-0781 & 49851.67410 & 0.00639 & Prim & B+R & MACHO \\
 OGLE-SMC-ECL-0781 & 49852.80913 & 0.00995 & Sec  & B+R & MACHO \\
 OGLE-SMC-ECL-0781 & 50201.46365 & 0.00429 & Prim & B+R & MACHO \\
 OGLE-SMC-ECL-0781 & 50202.62334 & 0.00207 & Sec  & B+R & MACHO \\
 OGLE-SMC-ECL-0781 & 50551.24174 & 0.00636 & Prim & B+R & MACHO \\
 OGLE-SMC-ECL-0781 & 50552.42087 & 0.00430 & Sec  & B+R & MACHO \\
 OGLE-SMC-ECL-0781 & 50851.52298 & 0.00449 & Prim & B+R & MACHO \\
 OGLE-SMC-ECL-0781 & 50852.72368 & 0.00320 & Sec  & B+R & MACHO \\
 OGLE-SMC-ECL-0781 & 51501.59099 & 0.00867 & Prim & B+R & MACHO \\
 OGLE-SMC-ECL-0781 & 51502.82123 & 0.00378 & Sec  & B+R & MACHO \\
 OGLE-SMC-ECL-0781 & 50950.51118 & 0.00632 & Prim & I   & OGLE II \\
 OGLE-SMC-ECL-0781 & 50951.72688 & 0.00346 & Sec  & I   & OGLE II \\
 OGLE-SMC-ECL-0781 & 51650.06399 & 0.00269 & Prim & I   & OGLE II \\
 OGLE-SMC-ECL-0781 & 51651.34170 & 0.00330 & Sec  & I   & OGLE II \\
 OGLE-SMC-ECL-0781 & 52201.12370 & 0.00245 & Prim & I   & OGLE III \\
 OGLE-SMC-ECL-0781 & 52202.45512 & 0.00330 & Sec  & I   & OGLE III \\
 OGLE-SMC-ECL-0781 & 52574.00067 & 0.00203 & Prim & I   & OGLE III \\
 OGLE-SMC-ECL-0781 & 52575.35665 & 0.00219 & Sec  & I   & OGLE III \\
 OGLE-SMC-ECL-0781 & 52923.77283 & 0.00191 & Prim & I   & OGLE III \\
 OGLE-SMC-ECL-0781 & 52925.16354 & 0.00283 & Sec  & I   & OGLE III \\
 OGLE-SMC-ECL-0781 & 53299.94961 & 0.00223 & Prim & I   & OGLE III \\
 OGLE-SMC-ECL-0781 & 53301.36089 & 0.00390 & Sec  & I   & OGLE III \\
 OGLE-SMC-ECL-0781 & 53649.71827 & 0.00569 & Prim & I   & OGLE III \\
 OGLE-SMC-ECL-0781 & 53651.16977 & 0.00287 & Sec  & I   & OGLE III \\
 OGLE-SMC-ECL-0781 & 53999.48473 & 0.00358 & Prim & I   & OGLE III \\
 OGLE-SMC-ECL-0781 & 54000.98147 & 0.00319 & Sec  & I   & OGLE III \\
 OGLE-SMC-ECL-0781 & 54349.25336 & 0.00150 & Prim & I   & OGLE III \\
 OGLE-SMC-ECL-0781 & 54350.78606 & 0.00575 & Sec  & I   & OGLE III \\
 OGLE-SMC-ECL-0781 & 54801.32027 & 0.00152 & Prim & I   & OGLE III \\
 OGLE-SMC-ECL-0781 & 54802.88593 & 0.00470 & Sec  & I   & OGLE III \\
 OGLE-SMC-ECL-0781 & 56637.71770 & 0.00052 & Sec  & R   & DK154 \\
 OGLE-SMC-ECL-0781 & 56675.55172 & 0.00225 & Prim & R   & DK154 \\
   \hline
 OGLE-SMC-ECL-0888 & 48650.12114 & 0.00495 & Prim & B+R & MACHO \\
 OGLE-SMC-ECL-0888 & 48651.22335 & 0.00508 & Sec  & B+R & MACHO \\
 OGLE-SMC-ECL-0888 & 49499.94006 & 0.00870 & Prim & B+R & MACHO \\
 OGLE-SMC-ECL-0888 & 49501.05324 & 0.00649 & Sec  & B+R & MACHO \\
 OGLE-SMC-ECL-0888 & 49849.07573 & 0.01044 & Prim & B+R & MACHO \\
 OGLE-SMC-ECL-0888 & 49850.20619 & 0.00930 & Sec  & B+R & MACHO \\
 OGLE-SMC-ECL-0888 & 50200.13984 & 0.00504 & Prim & B+R & MACHO \\
 OGLE-SMC-ECL-0888 & 50201.26861 & 0.00575 & Sec  & B+R & MACHO \\
 OGLE-SMC-ECL-0888 & 50549.27599 & 0.00811 & Prim & B+R & MACHO \\
  \hline
  \noalign{\smallskip}\hline
\end{tabular}
\end{minipage}
\end{table}

\begin{table}
 \centering
  \begin{minipage}{95mm}
 \fontsize{1.8mm}{2.4mm}\selectfont
 \caption{List of the minima timings used for the analysis.}
\begin{tabular}{ccclcl}
\hline\hline\noalign{\smallskip}
 Star       &    JD Hel.- &  Error & Type   &  Filter  & Source /     \\
            &   2400000   &  [day] &        &          & Observatory  \\
\noalign{\smallskip}\hline \noalign{\smallskip}
 OGLE-SMC-ECL-0888 & 50550.40710 & 0.01648 & Sec  & B+R & MACHO \\
 OGLE-SMC-ECL-0888 & 50850.46408 & 0.00330 & Prim & B+R & MACHO \\
 OGLE-SMC-ECL-0888 & 50851.59112 & 0.01159 & Sec  & B+R & MACHO \\
 OGLE-SMC-ECL-0888 & 51500.78388 & 0.00681 & Prim & B+R & MACHO \\
 OGLE-SMC-ECL-0888 & 51501.89603 & 0.00505 & Sec  & B+R & MACHO \\
 OGLE-SMC-ECL-0888 & 52199.09028 & 0.00528 & Prim &  I  & OGLE III \\
 OGLE-SMC-ECL-0888 & 52200.13742 & 0.00884 & Sec  &  I  & OGLE III \\
 OGLE-SMC-ECL-0888 & 52600.02846 & 0.00631 & Prim &  I  & OGLE III \\
 OGLE-SMC-ECL-0888 & 52601.06977 & 0.00736 & Sec  &  I  & OGLE III \\
 OGLE-SMC-ECL-0888 & 52949.18978 & 0.00333 & Prim &  I  & OGLE III \\
 OGLE-SMC-ECL-0888 & 52950.19052 & 0.00191 & Sec  &  I  & OGLE III \\
 OGLE-SMC-ECL-0888 & 53300.25949 & 0.00116 & Prim &  I  & OGLE III \\
 OGLE-SMC-ECL-0888 & 53301.23084 & 0.00399 & Sec  &  I  & OGLE III \\
 OGLE-SMC-ECL-0888 & 53674.34361 & 0.00409 & Prim &  I  & OGLE III \\
 OGLE-SMC-ECL-0888 & 53675.28319 & 0.01150 & Sec  &  I  & OGLE III \\
 OGLE-SMC-ECL-0888 & 54050.35279 & 0.00313 & Prim &  I  & OGLE III \\
 OGLE-SMC-ECL-0888 & 54051.26885 & 0.00038 & Sec  &  I  & OGLE III \\
 OGLE-SMC-ECL-0888 & 54374.55955 & 0.00380 & Prim &  I  & OGLE III \\
 OGLE-SMC-ECL-0888 & 54375.44152 & 0.00309 & Sec  &  I  & OGLE III \\
 OGLE-SMC-ECL-0888 & 54800.44108 & 0.00414 & Prim &  I  & OGLE III \\
 OGLE-SMC-ECL-0888 & 54801.28789 & 0.00905 & Sec  &  I  & OGLE III \\
 OGLE-SMC-ECL-0888 & 56648.59424 & 0.00055 & Sec  &  R  & DK154 \\
 OGLE-SMC-ECL-0888 & 56699.59546 & 0.00175 & Prim &  R  & DK154 \\
  \hline
 OGLE-SMC-ECL-1001 & 48850.21555 & 0.00273 & Prim & B+R & MACHO \\
 OGLE-SMC-ECL-1001 & 48850.79998 & 0.00609 & Sec  & B+R & MACHO \\
 OGLE-SMC-ECL-1001 & 49849.63212 & 0.00151 & Prim & B+R & MACHO \\
 OGLE-SMC-ECL-1001 & 49850.23064 & 0.00839 & Sec  & B+R & MACHO \\
 OGLE-SMC-ECL-1001 & 50199.41863 & 0.00345 & Prim & B+R & MACHO \\
 OGLE-SMC-ECL-1001 & 50200.02376 & 0.00438 & Sec  & B+R & MACHO \\
 OGLE-SMC-ECL-1001 & 50550.36477 & 0.00404 & Prim & B+R & MACHO \\
 OGLE-SMC-ECL-1001 & 50550.98731 & 0.00361 & Sec  & B+R & MACHO \\
 OGLE-SMC-ECL-1001 & 50850.18196 & 0.00182 & Prim & B+R & MACHO \\
 OGLE-SMC-ECL-1001 & 50850.81887 & 0.00465 & Sec  & B+R & MACHO \\
 OGLE-SMC-ECL-1001 & 51499.80461 & 0.00227 & Prim & B+R & MACHO \\
 OGLE-SMC-ECL-1001 & 51500.44253 & 0.00341 & Sec  & B+R & MACHO \\
 OGLE-SMC-ECL-1001 & 52199.40475 & 0.00254 & Prim &  I  & OGLE III \\
 OGLE-SMC-ECL-1001 & 52200.02416 & 0.00093 & Sec  &  I  & OGLE III \\
 OGLE-SMC-ECL-1001 & 52199.40475 & 0.00254 & Prim &  I  & OGLE III \\
 OGLE-SMC-ECL-1001 & 52575.37135 & 0.00282 & Sec  &  I  & OGLE III \\
 OGLE-SMC-ECL-1001 & 52924.58209 & 0.00070 & Prim &  I  & OGLE III \\
 OGLE-SMC-ECL-1001 & 52925.16579 & 0.00325 & Sec  &  I  & OGLE III \\
 OGLE-SMC-ECL-1001 & 53299.95187 & 0.00042 & Prim &  I  & OGLE III \\
 OGLE-SMC-ECL-1001 & 53300.52307 & 0.00219 & Sec  &  I  & OGLE III \\
 OGLE-SMC-ECL-1001 & 53649.74110 & 0.00006 & Prim &  I  & OGLE III \\
 OGLE-SMC-ECL-1001 & 53650.30821 & 0.00044 & Sec  &  I  & OGLE III \\
 OGLE-SMC-ECL-1001 & 53999.55700 & 0.00150 & Prim &  I  & OGLE III \\
 OGLE-SMC-ECL-1001 & 54000.08782 & 0.00011 & Sec  &  I  & OGLE III \\
 OGLE-SMC-ECL-1001 & 54350.51940 & 0.00014 & Prim &  I  & OGLE III \\
 OGLE-SMC-ECL-1001 & 54351.05432 & 0.00070 & Sec  &  I  & OGLE III \\
 OGLE-SMC-ECL-1001 & 54800.26261 & 0.00178 & Prim &  I  & OGLE III \\
 OGLE-SMC-ECL-1001 & 54800.78059 & 0.00223 & Sec  &  I  & OGLE III \\
 OGLE-SMC-ECL-1001 & 56641.60734 & 0.00122 & Sec  &  R  & DK154    \\
 OGLE-SMC-ECL-1001 & 56673.54658 & 0.00453 & Prim &  R  & DK154    \\
 OGLE-SMC-ECL-1001 & 56702.60206 & 0.00200 & Prim &  R  & DK154    \\
  \hline
 OGLE-SMC-ECL-1298 & 48700.80975 & 0.00568 & Prim & B+R & MACHO    \\
 OGLE-SMC-ECL-1298 & 48701.85571 & 0.00730 & Sec  & B+R & MACHO    \\
 OGLE-SMC-ECL-1298 & 49549.34310 & 0.00605 & Prim & B+R & MACHO    \\
 OGLE-SMC-ECL-1298 & 49550.42315 & 0.00753 & Sec  & B+R & MACHO    \\
 OGLE-SMC-ECL-1298 & 49849.13107 & 0.01288 & Prim & B+R & MACHO    \\
 OGLE-SMC-ECL-1298 & 49850.24535 & 0.00998 & Sec  & B+R & MACHO    \\
 OGLE-SMC-ECL-1298 & 50199.77445 & 0.00300 & Prim & B+R & MACHO    \\
 OGLE-SMC-ECL-1298 & 50200.89616 & 0.00724 & Sec  & B+R & MACHO    \\
 OGLE-SMC-ECL-1298 & 50550.42479 & 0.00531 & Prim & B+R & MACHO    \\
 OGLE-SMC-ECL-1298 & 50551.54191 & 0.00447 & Sec  & B+R & MACHO    \\
 OGLE-SMC-ECL-1298 & 50850.23337 & 0.00688 & Prim & B+R & MACHO    \\
 OGLE-SMC-ECL-1298 & 50851.33824 & 0.01225 & Sec  & B+R & MACHO    \\
 OGLE-SMC-ECL-1298 & 51500.69514 & 0.00657 & Prim & B+R & MACHO    \\
 OGLE-SMC-ECL-1298 & 51501.76815 & 0.01206 & Sec  & B+R & MACHO    \\
 OGLE-SMC-ECL-1298 & 50749.64318 & 0.00432 & Sec  &  I  & OGLE II  \\
 OGLE-SMC-ECL-1298 & 50750.29216 & 0.00168 & Prim &  I  & OGLE II  \\
 OGLE-SMC-ECL-1298 & 51360.43853 & 0.00065 & Prim &  I  & OGLE II  \\
 OGLE-SMC-ECL-1298 & 51361.49925 & 0.01471 & Sec  &  I  & OGLE II  \\
 OGLE-SMC-ECL-1298 & 52200.28714 & 0.00109 & Prim &  I  & OGLE III \\
 OGLE-SMC-ECL-1298 & 52201.22077 & 0.00300 & Sec  &  I  & OGLE III \\
 OGLE-SMC-ECL-1298 & 52600.04481 & 0.00253 & Prim &  I  & OGLE III \\
 OGLE-SMC-ECL-1298 & 52600.88875 & 0.00701 & Sec  &  I  & OGLE III \\
 OGLE-SMC-ECL-1298 & 52950.72224 & 0.00129 & Prim &  I  & OGLE III \\
 OGLE-SMC-ECL-1298 & 52951.52767 & 0.00227 & Sec  &  I  & OGLE III \\
 OGLE-SMC-ECL-1298 & 53299.62553 & 0.00384 & Prim &  I  & OGLE III \\
 OGLE-SMC-ECL-1298 & 53300.37165 & 0.03122 & Sec  &  I  & OGLE III \\
 OGLE-SMC-ECL-1298 & 53674.84148 & 0.00361 & Prim &  I  & OGLE III \\
  \hline
  \noalign{\smallskip}\hline
\end{tabular}
\end{minipage}
\end{table}

\begin{table}
 \centering
  \begin{minipage}{95mm}
 \fontsize{1.8mm}{2.4mm}\selectfont
 \caption{List of the minima timings used for the analysis.}
\begin{tabular}{ccclcl}
\hline\hline\noalign{\smallskip}
 Star       &    JD Hel.- &  Error & Type   &  Filter  & Source /     \\
            &   2400000   &  [day] &        &          & Observatory  \\
\noalign{\smallskip}\hline \noalign{\smallskip}
 OGLE-SMC-ECL-1298 & 53675.54332 & 0.00274 & Sec  &  I  & OGLE III \\
 OGLE-SMC-ECL-1298 & 54050.04304 & 0.00084 & Prim &  I  & OGLE III \\
 OGLE-SMC-ECL-1298 & 54050.71048 & 0.08333 & Sec  &  I  & OGLE III \\
 OGLE-SMC-ECL-1298 & 54374.40371 & 0.00178 & Prim &  I  & OGLE III \\
 OGLE-SMC-ECL-1298 & 54375.05651 & 0.00324 & Sec  &  I  & OGLE III \\
 OGLE-SMC-ECL-1298 & 54800.44849 & 0.00472 & Prim &  I  & OGLE III \\
 OGLE-SMC-ECL-1298 & 54801.08349 & 0.09458 & Sec  &  I  & OGLE III \\
 OGLE-SMC-ECL-1298 & 56397.53405 & 0.00545 & Prim &  R  & DK154    \\
 OGLE-SMC-ECL-1298 & 56580.67593 & 0.01310 & Sec  &  R  & DK154    \\
 OGLE-SMC-ECL-1298 & 56608.70669 & 0.00287 & Sec  &  R  & DK154    \\
 OGLE-SMC-ECL-1298 & 56673.62977 & 0.00118 & Sec  &  R  & DK154    \\
 OGLE-SMC-ECL-1298 & 56702.56995 & 0.00786 & Prim &  R  & DK154    \\
   \hline
 OGLE-SMC-ECL-1407 & 48649.17421 & 0.00323 & Prim & B+R & MACHO    \\
 OGLE-SMC-ECL-1407 & 48650.40889 & 0.00669 & Sec  & B+R & MACHO    \\
 OGLE-SMC-ECL-1407 & 49499.99436 & 0.00697 & Prim & B+R & MACHO    \\
 OGLE-SMC-ECL-1407 & 49501.21343 & 0.00811 & Sec  & B+R & MACHO    \\
 OGLE-SMC-ECL-1407 & 49850.83227 & 0.00611 & Prim & B+R & MACHO    \\
 OGLE-SMC-ECL-1407 & 49852.02643 & 0.00710 & Sec  & B+R & MACHO    \\
 OGLE-SMC-ECL-1407 & 50199.56787 & 0.01047 & Prim & B+R & MACHO    \\
 OGLE-SMC-ECL-1407 & 50200.74580 & 0.00719 & Sec  & B+R & MACHO    \\
 OGLE-SMC-ECL-1407 & 50550.39800 & 0.00301 & Prim & B+R & MACHO    \\
 OGLE-SMC-ECL-1407 & 50551.56030 & 0.00742 & Sec  & B+R & MACHO    \\
 OGLE-SMC-ECL-1407 & 50850.81538 & 0.00361 & Prim & B+R & MACHO    \\
 OGLE-SMC-ECL-1407 & 50851.95315 & 0.00580 & Sec  & B+R & MACHO    \\
 OGLE-SMC-ECL-1407 & 51499.96102 & 0.00528 & Prim & B+R & MACHO    \\
 OGLE-SMC-ECL-1407 & 51501.07678 & 0.01308 & Sec  & B+R & MACHO    \\
 OGLE-SMC-ECL-1407 & 52199.54100 & 0.00251 & Prim &  I  & OGLE III \\
 OGLE-SMC-ECL-1407 & 52200.59766 & 0.00496 & Sec  &  I  & OGLE III \\
 OGLE-SMC-ECL-1407 & 52575.58581 & 0.00118 & Prim &  I  & OGLE III \\
 OGLE-SMC-ECL-1407 & 52576.61588 & 0.00614 & Sec  &  I  & OGLE III \\
 OGLE-SMC-ECL-1407 & 52924.32346 & 0.00232 & Prim &  I  & OGLE III \\
 OGLE-SMC-ECL-1407 & 52925.31293 & 0.00140 & Sec  &  I  & OGLE III \\
 OGLE-SMC-ECL-1407 & 53300.37128 & 0.00145 & Prim &  I  & OGLE III \\
 OGLE-SMC-ECL-1407 & 53301.33887 & 0.00320 & Sec  &  I  & OGLE III \\
 OGLE-SMC-ECL-1407 & 53649.10211 & 0.00177 & Prim &  I  & OGLE III \\
 OGLE-SMC-ECL-1407 & 53650.06528 & 0.00404 & Sec  &  I  & OGLE III \\
 OGLE-SMC-ECL-1407 & 53999.94323 & 0.00416 & Prim &  I  & OGLE III \\
 OGLE-SMC-ECL-1407 & 54000.85550 & 0.00555 & Sec  &  I  & OGLE III \\
 OGLE-SMC-ECL-1407 & 54350.77939 & 0.00263 & Prim &  I  & OGLE III \\
 OGLE-SMC-ECL-1407 & 54351.69302 & 0.00463 & Sec  &  I  & OGLE III \\
 OGLE-SMC-ECL-1407 & 54800.35016 & 0.00327 & Prim &  I  & OGLE III \\
 OGLE-SMC-ECL-1407 & 54801.23674 & 0.00381 & Sec  &  I  & OGLE III \\
 OGLE-SMC-ECL-1407 & 56640.64137 & 0.00119 & Prim &  R  & DK154    \\
 OGLE-SMC-ECL-1407 & 56706.62008 & 0.00212 & Sec  &  R  & DK154    \\
  \hline
 OGLE-SMC-ECL-2186 & 48701.09378 & 0.00523 & Prim & B+R & MACHO    \\
 OGLE-SMC-ECL-2186 & 48702.98665 & 0.01720 & Sec  & B+R & MACHO    \\
 OGLE-SMC-ECL-2186 & 49550.25484 & 0.00192 & Prim & B+R & MACHO    \\
 OGLE-SMC-ECL-2186 & 49552.13771 & 0.01594 & Sec  & B+R & MACHO    \\
 OGLE-SMC-ECL-2186 & 49849.77123 & 0.00794 & Prim & B+R & MACHO    \\
 OGLE-SMC-ECL-2186 & 49851.64007 & 0.02195 & Sec  & B+R & MACHO    \\
 OGLE-SMC-ECL-2186 & 50198.66440 & 0.00176 & Prim & B+R & MACHO    \\
 OGLE-SMC-ECL-2186 & 50200.50302 & 0.00638 & Sec  & B+R & MACHO    \\
 OGLE-SMC-ECL-2186 & 50550.84602 & 0.00851 & Prim & B+R & MACHO    \\
 OGLE-SMC-ECL-2186 & 50552.66865 & 0.01982 & Sec  & B+R & MACHO    \\
 OGLE-SMC-ECL-2186 & 50850.36453 & 0.01027 & Prim & B+R & MACHO    \\
 OGLE-SMC-ECL-2186 & 50852.16776 & 0.00809 & Sec  & B+R & MACHO    \\
 OGLE-SMC-ECL-2186 & 51498.76490 & 0.00206 & Prim & B+R & MACHO    \\
 OGLE-SMC-ECL-2186 & 51500.54160 & 0.00316 & Sec  & B+R & MACHO    \\
 OGLE-SMC-ECL-2186 & 50748.32719 & 0.00435 & Prim &  I  & OGLE III \\
 OGLE-SMC-ECL-2186 & 50750.11970 & 0.00500 & Sec  &  I  & OGLE III \\
 OGLE-SMC-ECL-2186 & 51699.53127 & 0.00179 & Prim &  I  & OGLE III \\
 OGLE-SMC-ECL-2186 & 51701.28957 & 0.00242 & Sec  &  I  & OGLE III \\
 OGLE-SMC-ECL-2186 & 52199.83143 & 0.00657 & Prim &  I  & OGLE III \\
 OGLE-SMC-ECL-2186 & 52201.54632 & 0.00741 & Sec  &  I  & OGLE III \\
 OGLE-SMC-ECL-2186 & 52575.06110 & 0.00450 & Prim &  I  & OGLE III \\
 OGLE-SMC-ECL-2186 & 52576.73683 & 0.00513 & Sec  &  I  & OGLE III \\
 OGLE-SMC-ECL-2186 & 52923.96294 & 0.00808 & Prim &  I  & OGLE III \\
 OGLE-SMC-ECL-2186 & 52925.60931 & 0.00509 & Sec  &  I  & OGLE III \\
 OGLE-SMC-ECL-2186 & 53299.17085 & 0.00203 & Prim &  I  & OGLE III \\
 OGLE-SMC-ECL-2186 & 53300.80791 & 0.00373 & Sec  &  I  & OGLE III \\
 OGLE-SMC-ECL-2186 & 53651.34727 & 0.00639 & Prim &  I  & OGLE III \\
 OGLE-SMC-ECL-2186 & 53652.96525 & 0.00485 & Sec  &  I  & OGLE III \\
 OGLE-SMC-ECL-2186 & 54000.22926 & 0.00325 & Prim &  I  & OGLE III \\
 OGLE-SMC-ECL-2186 & 54001.83984 & 0.00567 & Sec  &  I  & OGLE III \\
 OGLE-SMC-ECL-2186 & 54349.12883 & 0.00566 & Prim &  I  & OGLE III \\
 OGLE-SMC-ECL-2186 & 54350.69183 & 0.00641 & Sec  &  I  & OGLE III \\
 OGLE-SMC-ECL-2186 & 54800.04630 & 0.00450 & Prim &  I  & OGLE III \\
 OGLE-SMC-ECL-2186 & 54801.58628 & 0.00743 & Sec  &  I  & OGLE III \\
 OGLE-SMC-ECL-2186 & 56669.56854 & 0.00091 & Prim &  R  & DK154    \\
 OGLE-SMC-ECL-2186 & 56677.58220 & 0.00066 & Sec  &  R  & DK154    \\
   \hline
 OGLE-SMC-ECL-2225 & 48699.87030 & 0.00966 & Prim & B+R & MACHO    \\
  \hline
  \noalign{\smallskip}\hline
\end{tabular}
\end{minipage}
\end{table}

\begin{table}
 \centering
  \begin{minipage}{95mm}
 \fontsize{1.8mm}{2.4mm}\selectfont
 \caption{List of the minima timings used for the analysis.}
\begin{tabular}{ccclcl}
\hline\hline\noalign{\smallskip}
 Star       &    JD Hel.- &  Error & Type   &  Filter  & Source /     \\
            &   2400000   &  [day] &        &          & Observatory  \\
\noalign{\smallskip}\hline \noalign{\smallskip}
 OGLE-SMC-ECL-2225 & 48700.45057 & 0.00607 & Sec  & B+R & MACHO    \\
 OGLE-SMC-ECL-2225 & 49550.14572 & 0.00366 & Prim & B+R & MACHO    \\
 OGLE-SMC-ECL-2225 & 49550.73392 & 0.00506 & Sec  & B+R & MACHO    \\
 OGLE-SMC-ECL-2225 & 49849.97432 & 0.00275 & Prim & B+R & MACHO    \\
 OGLE-SMC-ECL-2225 & 49850.58771 & 0.00696 & Sec  & B+R & MACHO    \\
 OGLE-SMC-ECL-2225 & 50200.52770 & 0.00631 & Prim & B+R & MACHO    \\
 OGLE-SMC-ECL-2225 & 50201.12738 & 0.00641 & Sec  & B+R & MACHO    \\
 OGLE-SMC-ECL-2225 & 50549.57873 & 0.01289 & Prim & B+R & MACHO    \\
 OGLE-SMC-ECL-2225 & 50550.20266 & 0.00979 & Sec  & B+R & MACHO    \\
 OGLE-SMC-ECL-2225 & 50849.39455 & 0.00747 & Prim & B+R & MACHO    \\
 OGLE-SMC-ECL-2225 & 50850.04827 & 0.00596 & Sec  & B+R & MACHO    \\
 OGLE-SMC-ECL-2225 & 51499.76771 & 0.02076 & Prim & B+R & MACHO    \\
 OGLE-SMC-ECL-2225 & 51500.44540 & 0.02053 & Sec  & B+R & MACHO    \\
 OGLE-SMC-ECL-2225 & 52199.34712 & 0.00252 & Prim &  I  & OGLE III \\
 OGLE-SMC-ECL-2225 & 52200.10630 & 0.01075 & Sec  &  I  & OGLE III \\
 OGLE-SMC-ECL-2225 & 52575.25680 & 0.00215 & Prim &  I  & OGLE III \\
 OGLE-SMC-ECL-2225 & 52576.02555 & 0.00544 & Sec  &  I  & OGLE III \\
 OGLE-SMC-ECL-2225 & 52924.29983 & 0.00319 & Prim &  I  & OGLE III \\
 OGLE-SMC-ECL-2225 & 52925.10071 & 0.00252 & Sec  &  I  & OGLE III \\
 OGLE-SMC-ECL-2225 & 53300.20601 & 0.00534 & Prim &  I  & OGLE III \\
 OGLE-SMC-ECL-2225 & 53301.02592 & 0.00359 & Sec  &  I  & OGLE III \\
 OGLE-SMC-ECL-2225 & 53649.24349 & 0.00271 & Prim &  I  & OGLE III \\
 OGLE-SMC-ECL-2225 & 53650.10421 & 0.00207 & Sec  &  I  & OGLE III \\
 OGLE-SMC-ECL-2225 & 53999.78851 & 0.00337 & Prim &  I  & OGLE III \\
 OGLE-SMC-ECL-2225 & 54000.67118 & 0.00923 & Sec  &  I  & OGLE III \\
 OGLE-SMC-ECL-2225 & 54350.35287 & 0.00291 & Prim &  I  & OGLE III \\
 OGLE-SMC-ECL-2225 & 54351.22518 & 0.00150 & Sec  &  I  & OGLE III \\
 OGLE-SMC-ECL-2225 & 54799.33928 & 0.00355 & Prim &  I  & OGLE III \\
 OGLE-SMC-ECL-2225 & 54800.24310 & 0.00304 & Sec  &  I  & OGLE III \\
 OGLE-SMC-ECL-2225 & 56641.62568 & 0.00099 & Prim &  R  & DK154    \\
 OGLE-SMC-ECL-2225 & 56677.43050 & 0.00213 & Prim &  R  & DK154    \\
 OGLE-SMC-ECL-2225 & 56706.66141 & 0.00932 & Sec  &  R  & DK154    \\
 OGLE-SMC-ECL-2225 & 56727.57026 & 0.00856 & Sec  &  R  & DK154    \\
 OGLE-SMC-ECL-2225 & 56739.49729 & 0.00448 & Sec  &  R  & DK154    \\
  \hline
 OGLE-SMC-ECL-2251 & 50750.22871 & 0.01111 & Prim &  I  & OGLE II   \\
 OGLE-SMC-ECL-2251 & 50751.76132 & 0.00909 & Sec  &  I  & OGLE II   \\
 OGLE-SMC-ECL-2251 & 51701.03998 & 0.00534 & Prim &  I  & OGLE II   \\
 OGLE-SMC-ECL-2251 & 51702.49569 & 0.00466 & Sec  &  I  & OGLE II   \\
 OGLE-SMC-ECL-2251 & 52200.98659 & 0.00419 & Prim &  I  & OGLE III  \\
 OGLE-SMC-ECL-2251 & 52202.35316 & 0.00760 & Sec  &  I  & OGLE III  \\
 OGLE-SMC-ECL-2251 & 52574.77804 & 0.00435 & Prim &  I  & OGLE III  \\
 OGLE-SMC-ECL-2251 & 52576.09912 & 0.01312 & Sec  &  I  & OGLE III  \\
 OGLE-SMC-ECL-2251 & 52925.18554 & 0.00564 & Prim &  I  & OGLE III  \\
 OGLE-SMC-ECL-2251 & 52926.45681 & 0.02329 & Sec  &  I  & OGLE III  \\
 OGLE-SMC-ECL-2251 & 53298.99286 & 0.00406 & Prim &  I  & OGLE III  \\
 OGLE-SMC-ECL-2251 & 53300.22628 & 0.02132 & Sec  &  I  & OGLE III  \\
 OGLE-SMC-ECL-2251 & 53649.40932 & 0.00442 & Prim &  I  & OGLE III  \\
 OGLE-SMC-ECL-2251 & 53650.58269 & 0.00382 & Sec  &  I  & OGLE III  \\
 OGLE-SMC-ECL-2251 & 53999.83046 & 0.00299 & Prim &  I  & OGLE III  \\
 OGLE-SMC-ECL-2251 & 54000.94196 & 0.00673 & Sec  &  I  & OGLE III  \\
 OGLE-SMC-ECL-2251 & 54350.26571 & 0.00468 & Prim &  I  & OGLE III  \\
 OGLE-SMC-ECL-2251 & 54351.32752 & 0.03865 & Sec  &  I  & OGLE III  \\
 OGLE-SMC-ECL-2251 & 54801.13489 & 0.00506 & Prim &  I  & OGLE III  \\
 OGLE-SMC-ECL-2251 & 54802.18335 & 0.01739 & Sec  &  I  & OGLE III  \\
 OGLE-SMC-ECL-2251 & 56639.69916 & 0.00113 & Prim &  R  & DK154     \\
 OGLE-SMC-ECL-2251 & 56675.56777 & 0.00287 & Sec  &  R  & DK154     \\
 OGLE-SMC-ECL-2251 & 56689.57815 & 0.00124 & Sec  &  R  & DK154     \\
  \hline
 OGLE-SMC-ECL-2524 & 48749.05117 & 0.00480 & Prim & B+R & MACHO \\
 OGLE-SMC-ECL-2524 & 48750.47954 & 0.00177 & Sec  & B+R & MACHO \\
 OGLE-SMC-ECL-2524 & 49749.07351 & 0.00122 & Prim & B+R & MACHO \\
 OGLE-SMC-ECL-2524 & 49750.50676 & 0.00382 & Sec  & B+R & MACHO \\
 OGLE-SMC-ECL-2524 & 50250.18009 & 0.00171 & Prim & B+R & MACHO \\
 OGLE-SMC-ECL-2524 & 50251.61017 & 0.00254 & Sec  & B+R & MACHO \\
 OGLE-SMC-ECL-2524 & 50749.11909 & 0.00208 & Prim & B+R & MACHO \\
 OGLE-SMC-ECL-2524 & 50750.53425 & 0.00654 & Sec  & B+R & MACHO \\
 OGLE-SMC-ECL-2524 & 51499.69730 & 0.00381 & Prim & B+R & MACHO \\
 OGLE-SMC-ECL-2524 & 51501.07663 & 0.00180 & Sec  & B+R & MACHO \\
 OGLE-SMC-ECL-2524 & 52200.40699 & 0.00311 & Prim &  I  & OGLE III \\
 OGLE-SMC-ECL-2524 & 52201.70247 & 0.00847 & Sec  &  I  & OGLE III \\
 OGLE-SMC-ECL-2524 & 52575.70108 & 0.00258 & Prim &  I  & OGLE III \\
 OGLE-SMC-ECL-2524 & 52576.96208 & 0.00417 & Sec  &  I  & OGLE III \\
 OGLE-SMC-ECL-2524 & 52924.96844 & 0.00138 & Prim &  I  & OGLE III \\
 OGLE-SMC-ECL-2524 & 52926.17872 & 0.00242 & Sec  &  I  & OGLE III \\
 OGLE-SMC-ECL-2524 & 53300.26282 & 0.00111 & Prim &  I  & OGLE III \\
 OGLE-SMC-ECL-2524 & 53301.42517 & 0.00349 & Sec  &  I  & OGLE III \\
 OGLE-SMC-ECL-2524 & 53649.53104 & 0.00122 & Prim &  I  & OGLE III \\
 OGLE-SMC-ECL-2524 & 53650.64581 & 0.00225 & Sec  &  I  & OGLE III \\
 OGLE-SMC-ECL-2524 & 54000.95733 & 0.00455 & Prim &  I  & OGLE III \\
 OGLE-SMC-ECL-2524 & 54002.02888 & 0.00174 & Sec  &  I  & OGLE III \\
 OGLE-SMC-ECL-2524 & 54350.23684 & 0.00791 & Prim &  I  & OGLE III \\
 OGLE-SMC-ECL-2524 & 54351.24487 & 0.00204 & Sec  &  I  & OGLE III \\
  \hline
  \noalign{\smallskip}\hline
\end{tabular}
\end{minipage}
\end{table}

\begin{table}
 \centering
  \begin{minipage}{95mm}
 \fontsize{1.8mm}{2.4mm}\selectfont
 \caption{List of the minima timings used for the analysis.}
\begin{tabular}{ccclcl}
\hline\hline\noalign{\smallskip}
 Star       &    JD Hel.- &  Error & Type   &  Filter  & Source /     \\
            &   2400000   &  [day] &        &          & Observatory  \\
\noalign{\smallskip}\hline \noalign{\smallskip}
 OGLE-SMC-ECL-2524 & 54799.30059 & 0.00479 & Prim &  I  & OGLE III \\
 OGLE-SMC-ECL-2524 & 54800.24413 & 0.00068 & Sec  &  I  & OGLE III \\
 OGLE-SMC-ECL-2524 & 56639.65255 & 0.00061 & Sec  &  R  & DK154 \\
 OGLE-SMC-ECL-2524 & 56673.59328 & 0.00059 & Prim &  R  & DK154 \\
 OGLE-SMC-ECL-2524 & 56689.54270 & 0.00350 & Sec  &  R  & DK154 \\
  \hline
 OGLE-SMC-ECL-2534 & 48800.83606 & 0.00410 & Prim & B+R & MACHO \\
 OGLE-SMC-ECL-2534 & 48801.88294 & 0.01385 & Sec  & B+R & MACHO \\
 OGLE-SMC-ECL-2534 & 49799.92570 & 0.00731 & Prim & B+R & MACHO \\
 OGLE-SMC-ECL-2534 & 49800.96534 & 0.00873 & Sec  & B+R & MACHO \\
 OGLE-SMC-ECL-2534 & 50199.53392 & 0.01529 & Prim & B+R & MACHO \\
 OGLE-SMC-ECL-2534 & 50200.59122 & 0.00513 & Sec  & B+R & MACHO \\
 OGLE-SMC-ECL-2534 & 50550.94672 & 0.01128 & Prim & B+R & MACHO \\
 OGLE-SMC-ECL-2534 & 50552.00100 & 0.01895 & Sec  & B+R & MACHO \\
 OGLE-SMC-ECL-2534 & 50849.50052 & 0.00706 & Prim & B+R & MACHO \\
 OGLE-SMC-ECL-2534 & 50850.58736 & 0.02162 & Sec  & B+R & MACHO \\
 OGLE-SMC-ECL-2534 & 51499.48078 & 0.02502 & Prim & B+R & MACHO \\
 OGLE-SMC-ECL-2534 & 51500.57635 & 0.01736 & Sec  & B+R & MACHO \\
 OGLE-SMC-ECL-2534 & 52199.97862 & 0.00262 & Prim &  I  & OGLE III \\
 OGLE-SMC-ECL-2534 & 52201.08392 & 0.00473 & Sec  &  I  & OGLE III \\
 OGLE-SMC-ECL-2534 & 52574.34158 & 0.00194 & Prim &  I  & OGLE III \\
 OGLE-SMC-ECL-2534 & 52575.45897 & 0.00952 & Sec  &  I  & OGLE III \\
 OGLE-SMC-ECL-2534 & 52925.72595 & 0.00185 & Prim &  I  & OGLE III \\
 OGLE-SMC-ECL-2534 & 52926.85598 & 0.00211 & Sec  &  I  & OGLE III \\
 OGLE-SMC-ECL-2534 & 53300.09757 & 0.00093 & Prim &  I  & OGLE III \\
 OGLE-SMC-ECL-2534 & 53301.23659 & 0.00496 & Sec  &  I  & OGLE III \\
 OGLE-SMC-ECL-2534 & 53649.19553 & 0.00172 & Prim &  I  & OGLE III \\
 OGLE-SMC-ECL-2534 & 53650.34464 & 0.00781 & Sec  &  I  & OGLE III \\
 OGLE-SMC-ECL-2534 & 54000.59273 & 0.00031 & Prim &  I  & OGLE III \\
 OGLE-SMC-ECL-2534 & 54001.74718 & 0.00343 & Sec  &  I  & OGLE III \\
 OGLE-SMC-ECL-2534 & 54349.68687 & 0.00154 & Prim &  I  & OGLE III \\
 OGLE-SMC-ECL-2534 & 54350.86174 & 0.00322 & Sec  &  I  & OGLE III \\
 OGLE-SMC-ECL-2534 & 54799.84807 & 0.00475 & Prim &  I  & OGLE III \\
 OGLE-SMC-ECL-2534 & 54801.03243 & 0.01016 & Sec  &  I  & OGLE III \\
  \hline
 OGLE-SMC-ECL-3594 & 48751.03362 & 0.00612 & Prim & B+R & MACHO \\
 OGLE-SMC-ECL-3594 & 48752.73313 & 0.00487 & Sec  & B+R & MACHO \\
 OGLE-SMC-ECL-3594 & 49651.74109 & 0.00235 & Prim & B+R & MACHO \\
 OGLE-SMC-ECL-3594 & 49653.46092 & 0.00449 & Sec  & B+R & MACHO \\
 OGLE-SMC-ECL-3594 & 49998.15595 & 0.00554 & Prim & B+R & MACHO \\
 OGLE-SMC-ECL-3594 & 49999.88282 & 0.00592 & Sec  & B+R & MACHO \\
 OGLE-SMC-ECL-3594 & 50348.91199 & 0.00310 & Prim & B+R & MACHO \\
 OGLE-SMC-ECL-3594 & 50350.65164 & 0.01405 & Sec  & B+R & MACHO \\
 OGLE-SMC-ECL-3594 & 50652.02791 & 0.00688 & Prim & B+R & MACHO \\
 OGLE-SMC-ECL-3594 & 50653.78149 & 0.00711 & Sec  & B+R & MACHO \\
 OGLE-SMC-ECL-3594 & 50950.81588 & 0.00496 & Prim & B+R & MACHO \\
 OGLE-SMC-ECL-3594 & 50952.56833 & 0.00282 & Sec  & B+R & MACHO \\
 OGLE-SMC-ECL-3594 & 51548.38929 & 0.00442 & Prim & B+R & MACHO \\
 OGLE-SMC-ECL-3594 & 51550.18670 & 0.00918 & Sec  & B+R & MACHO \\
 OGLE-SMC-ECL-3594 & 52202.24613 & 0.00398 & Prim &  I  & OGLE III \\
 OGLE-SMC-ECL-3594 & 52204.07478 & 0.00003 & Sec  &  I  & OGLE III \\
 OGLE-SMC-ECL-3594 & 52574.64781 & 0.00245 & Prim &  I  & OGLE III \\
 OGLE-SMC-ECL-3594 & 52576.49461 & 0.00434 & Sec  &  I  & OGLE III \\
 OGLE-SMC-ECL-3594 & 52925.39286 & 0.00217 & Prim &  I  & OGLE III \\
 OGLE-SMC-ECL-3594 & 52927.26649 & 0.00003 & Sec  &  I  & OGLE III \\
 OGLE-SMC-ECL-3594 & 53302.12389 & 0.00793 & Prim &  I  & OGLE III \\
 OGLE-SMC-ECL-3594 & 53304.00261 & 0.00601 & Sec  &  I  & OGLE III \\
 OGLE-SMC-ECL-3594 & 53648.54075 & 0.00042 & Prim &  I  & OGLE III \\
 OGLE-SMC-ECL-3594 & 53650.43839 & 0.00270 & Sec  &  I  & OGLE III \\
 OGLE-SMC-ECL-3594 & 53999.28473 & 0.00329 & Prim &  I  & OGLE III \\
 OGLE-SMC-ECL-3594 & 54001.21997 & 0.00223 & Sec  &  I  & OGLE III \\
 OGLE-SMC-ECL-3594 & 54350.03062 & 0.00353 & Prim &  I  & OGLE III \\
 OGLE-SMC-ECL-3594 & 54351.97934 & 0.00245 & Sec  &  I  & OGLE III \\
 OGLE-SMC-ECL-3594 & 54800.37520 & 0.00173 & Prim &  I  & OGLE III \\
 OGLE-SMC-ECL-3594 & 54802.34160 & 0.00842 & Sec  &  I  & OGLE III \\
  \hline
 OGLE-SMC-ECL-3677 & 48699.80729 & 0.01272 & Sec  & B+R  & MACHO \\
 OGLE-SMC-ECL-3677 & 48701.90985 & 0.01951 & Prim & B+R  & MACHO \\
 OGLE-SMC-ECL-3677 & 49548.92971 & 0.01937 & Sec  & B+R  & MACHO \\
 OGLE-SMC-ECL-3677 & 49551.04309 & 0.00854 & Prim & B+R  & MACHO \\
 OGLE-SMC-ECL-3677 & 49847.71464 & 0.02852 & Sec  & B+R  & MACHO \\
 OGLE-SMC-ECL-3677 & 49849.81264 & 0.01114 & Prim & B+R  & MACHO \\
 OGLE-SMC-ECL-3677 & 50198.88911 & 0.01522 & Sec  & B+R  & MACHO \\
 OGLE-SMC-ECL-3677 & 50201.00269 & 0.01166 & Prim & B+R  & MACHO \\
 OGLE-SMC-ECL-3677 & 50550.07712 & 0.02286 & Sec  & B+R  & MACHO \\
 OGLE-SMC-ECL-3677 & 50552.19617 & 0.01038 & Prim & B+R  & MACHO \\
 OGLE-SMC-ECL-3677 & 50848.82438 & 0.02664 & Sec  & B+R  & MACHO \\
 OGLE-SMC-ECL-3677 & 50850.97001 & 0.00644 & Prim & B+R  & MACHO \\
 OGLE-SMC-ECL-3677 & 51498.77983 & 0.02237 & Sec  & B+R  & MACHO \\
 OGLE-SMC-ECL-3677 & 51500.92187 & 0.00895 & Prim & B+R  & MACHO \\
 OGLE-SMC-ECL-3677 & 52201.14249 & 0.01502 & Sec  &  I   & OGLE III \\
 OGLE-SMC-ECL-3677 & 52203.31297 & 0.00469 & Prim &  I   & OGLE III \\
 OGLE-SMC-ECL-3677 & 52573.27233 & 0.00763 & Sec  &  I   & OGLE III \\
 OGLE-SMC-ECL-3677 & 52575.46868 & 0.00569 & Prim &  I   & OGLE III \\
  \hline
  \noalign{\smallskip}\hline
\end{tabular}
\end{minipage}
\end{table}

\begin{table}
 \centering
  \begin{minipage}{95mm}
 \fontsize{1.8mm}{2.4mm}\selectfont
 \caption{List of the minima timings used for the analysis.}
\begin{tabular}{ccclcl}
\hline\hline\noalign{\smallskip}
 Star       &    JD Hel.- &  Error & Type   &  Filter  & Source /     \\
            &   2400000   &  [day] &        &          & Observatory  \\
\noalign{\smallskip}\hline \noalign{\smallskip}
 OGLE-SMC-ECL-3677 & 52924.45135 & 0.00696 & Sec  &  I   & OGLE III \\
 OGLE-SMC-ECL-3677 & 52926.65837 & 0.00648 & Prim &  I   & OGLE III \\
 OGLE-SMC-ECL-3677 & 53301.81466 & 0.01272 & Sec  &  I   & OGLE III \\
 OGLE-SMC-ECL-3677 & 53304.07030 & 0.00956 & Prim &  I   & OGLE III \\
 OGLE-SMC-ECL-3677 & 53647.75372 & 0.01490 & Sec  &  I   & OGLE III \\
 OGLE-SMC-ECL-3677 & 53650.01979 & 0.00356 & Prim &  I   & OGLE III \\
 OGLE-SMC-ECL-3677 & 53998.92989 & 0.01257 & Sec  &  I   & OGLE III \\
 OGLE-SMC-ECL-3677 & 54001.21811 & 0.00470 & Prim &  I   & OGLE III \\
 OGLE-SMC-ECL-3677 & 54350.11175 & 0.01169 & Sec  &  I   & OGLE III \\
 OGLE-SMC-ECL-3677 & 54352.39886 & 0.00198 & Prim &  I   & OGLE III \\
 OGLE-SMC-ECL-3677 & 54800.87834 & 0.00487 & Sec  &  I   & OGLE III \\
 OGLE-SMC-ECL-3677 & 54803.18740 & 0.00587 & Prim &  I   & OGLE III \\
  \hline
 OGLE-SMC-ECL-3951 & 50548.54640 & 0.00520 & Prim &  I  & OGLE II \\
 OGLE-SMC-ECL-3951 & 50550.20977 & 0.00469 & Sec  &  I  & OGLE II \\
 OGLE-SMC-ECL-3951 & 51101.11593 & 0.00465 & Prim &  I  & OGLE II \\
 OGLE-SMC-ECL-3951 & 51102.74425 & 0.00397 & Sec  &  I  & OGLE II \\
 OGLE-SMC-ECL-3951 & 51700.25361 & 0.00386 & Prim &  I  & OGLE II \\
 OGLE-SMC-ECL-3951 & 51701.88334 & 0.00627 & Sec  &  I  & OGLE II \\
 OGLE-SMC-ECL-3951 & 52200.06093 & 0.00382 & Prim &  I  & OGLE III \\
 OGLE-SMC-ECL-3951 & 52201.66655 & 0.00436 & Sec  &  I  & OGLE III \\
 OGLE-SMC-ECL-3951 & 52575.68577 & 0.00775 & Prim &  I  & OGLE III \\
 OGLE-SMC-ECL-3951 & 52577.27291 & 0.00530 & Sec  &  I  & OGLE III \\
 OGLE-SMC-ECL-3951 & 52926.47673 & 0.00762 & Prim &  I  & OGLE III \\
 OGLE-SMC-ECL-3951 & 52928.03171 & 0.00860 & Sec  &  I  & OGLE III \\
 OGLE-SMC-ECL-3951 & 53298.99469 & 0.00262 & Prim &  I  & OGLE III \\
 OGLE-SMC-ECL-3951 & 53300.53536 & 0.00247 & Sec  &  I  & OGLE III \\
 OGLE-SMC-ECL-3951 & 53649.79107 & 0.00178 & Prim &  I  & OGLE III \\
 OGLE-SMC-ECL-3951 & 53651.30663 & 0.00770 & Sec  &  I  & OGLE III \\
 OGLE-SMC-ECL-3951 & 54000.58489 & 0.00707 & Prim &  I  & OGLE III \\
 OGLE-SMC-ECL-3951 & 54002.10747 & 0.00979 & Sec  &  I  & OGLE III \\
 OGLE-SMC-ECL-3951 & 54351.38258 & 0.00162 & Prim &  I  & OGLE III \\
 OGLE-SMC-ECL-3951 & 54352.86482 & 0.00654 & Sec  &  I  & OGLE III \\
 OGLE-SMC-ECL-3951 & 54801.52124 & 0.00479 & Prim &  I  & OGLE III \\
 OGLE-SMC-ECL-3951 & 54802.98917 & 0.01250 & Sec  &  I  & OGLE III \\
  \hline
 OGLE-SMC-ECL-4955 & 48750.12518 & 0.00631 & Prim & B+R & MACHO \\
 OGLE-SMC-ECL-4955 & 48751.38801 & 0.02229 & Sec  & B+R & MACHO \\
 OGLE-SMC-ECL-4955 & 49651.08861 & 0.00532 & Prim & B+R & MACHO \\
 OGLE-SMC-ECL-4955 & 49652.33413 & 0.03385 & Sec  & B+R & MACHO \\
 OGLE-SMC-ECL-4955 & 50000.39638 & 0.00395 & Prim & B+R & MACHO \\
 OGLE-SMC-ECL-4955 & 50001.62993 & 0.02874 & Sec  & B+R & MACHO \\
 OGLE-SMC-ECL-4955 & 50349.71256 & 0.00389 & Prim & B+R & MACHO \\
 OGLE-SMC-ECL-4955 & 50350.87651 & 0.01422 & Sec  & B+R & MACHO \\
 OGLE-SMC-ECL-4955 & 50649.11491 & 0.00313 & Prim & B+R & MACHO \\
 OGLE-SMC-ECL-4955 & 50650.25954 & 0.01288 & Sec  & B+R & MACHO \\
 OGLE-SMC-ECL-4955 & 50951.29501 & 0.00429 & Prim & B+R & MACHO \\
 OGLE-SMC-ECL-4955 & 50952.44016 & 0.01871 & Sec  & B+R & MACHO \\
 OGLE-SMC-ECL-4955 & 51550.09921 & 0.00687 & Prim & B+R & MACHO \\
 OGLE-SMC-ECL-4955 & 51551.18926 & 0.01608 & Sec  & B+R & MACHO \\
 OGLE-SMC-ECL-4955 & 52198.83073 & 0.00972 & Prim &  I  & OGLE III \\
 OGLE-SMC-ECL-4955 & 52199.82288 & 0.00238 & Sec  &  I  & OGLE III \\
 OGLE-SMC-ECL-4955 & 52575.85660 & 0.00253 & Prim &  I  & OGLE III \\
 OGLE-SMC-ECL-4955 & 52576.84703 & 0.00743 & Sec  &  I  & OGLE III \\
 OGLE-SMC-ECL-4955 & 52925.15650 & 0.00209 & Prim &  I  & OGLE III \\
 OGLE-SMC-ECL-4955 & 52926.11742 & 0.00000 & Sec  &  I  & OGLE III \\
 OGLE-SMC-ECL-4955 & 53299.40496 & 0.00020 & Prim &  I  & OGLE III \\
 OGLE-SMC-ECL-4955 & 53300.35027 & 0.00720 & Sec  &  I  & OGLE III \\
 OGLE-SMC-ECL-4955 & 53651.48950 & 0.00772 & Prim &  I  & OGLE III \\
 OGLE-SMC-ECL-4955 & 53652.42259 & 0.00203 & Sec  &  I  & OGLE III \\
 OGLE-SMC-ECL-4955 & 54000.79736 & 0.00435 & Prim &  I  & OGLE III \\
 OGLE-SMC-ECL-4955 & 54001.70010 & 0.00418 & Sec  &  I  & OGLE III \\
 OGLE-SMC-ECL-4955 & 54350.10388 & 0.00792 & Prim &  I  & OGLE III \\
 OGLE-SMC-ECL-4955 & 54350.99198 & 0.00293 & Sec  &  I  & OGLE III \\
 OGLE-SMC-ECL-4955 & 54799.20041 & 0.00141 & Prim &  I  & OGLE III \\
 OGLE-SMC-ECL-4955 & 54800.08989 & 0.00558 & Sec  &  I  & OGLE III \\
  \hline
 OGLE-SMC-ECL-5233 & 52198.73233 & 0.00798 & Prim &  I  & OGLE III \\
 OGLE-SMC-ECL-5233 & 52201.92612 & 0.00000 & Sec  &  I  & OGLE III \\
 OGLE-SMC-ECL-5233 & 52573.81356 & 0.00377 & Prim &  I  & OGLE III \\
 OGLE-SMC-ECL-5233 & 52576.99968 & 0.00250 & Sec  &  I  & OGLE III \\
 OGLE-SMC-ECL-5233 & 52923.55994 & 0.01665 & Prim &  I  & OGLE III \\
 OGLE-SMC-ECL-5233 & 52926.74636 & 0.00955 & Sec  &  I  & OGLE III \\
 OGLE-SMC-ECL-5233 & 53298.63156 & 0.00478 & Prim &  I  & OGLE III \\
 OGLE-SMC-ECL-5233 & 53301.76315 & 0.00548 & Sec  &  I  & OGLE III \\
 OGLE-SMC-ECL-5233 & 53648.41252 & 0.03105 & Prim &  I  & OGLE III \\
 OGLE-SMC-ECL-5233 & 53651.42894 & 0.00478 & Sec  &  I  & OGLE III \\
 OGLE-SMC-ECL-5233 & 53998.17633 & 0.01331 & Prim &  I  & OGLE III \\
 OGLE-SMC-ECL-5233 & 54001.10599 & 0.00849 & Sec  &  I  & OGLE III \\
 OGLE-SMC-ECL-5233 & 54347.97205 & 0.00515 & Prim &  I  & OGLE III \\
 OGLE-SMC-ECL-5233 & 54350.74730 & 0.01452 & Sec  &  I  & OGLE III \\
 OGLE-SMC-ECL-5233 & 54799.11105 & 0.02745 & Prim &  I  & OGLE III \\
 OGLE-SMC-ECL-5233 & 54801.71854 & 0.00803 & Sec  &  I  & OGLE III \\
 OGLE-SMC-ECL-5233 & 48747.75771 & 0.02284 & Prim & B+R & MACHO    \\
  \hline
  \noalign{\smallskip}\hline
\end{tabular}
\end{minipage}
\end{table}

\begin{table}
 \centering
  \begin{minipage}{95mm}
 \fontsize{1.8mm}{2.4mm}\selectfont
 \caption{List of the minima timings used for the analysis.}
\begin{tabular}{ccclcl}
\hline\hline\noalign{\smallskip}
 Star       &    JD Hel.- &  Error & Type   &  Filter  & Source /     \\
            &   2400000   &  [day] &        &          & Observatory  \\
\noalign{\smallskip}\hline \noalign{\smallskip}
 OGLE-SMC-ECL-5233 & 48749.89493 & 0.01255 & Sec  & B+R & MACHO    \\
 OGLE-SMC-ECL-5233 & 49751.21107 & 0.01683 & Prim & B+R & MACHO    \\
 OGLE-SMC-ECL-5233 & 49753.48924 & 0.01297 & Sec  & B+R & MACHO    \\
 OGLE-SMC-ECL-5233 & 50247.75172 & 0.03015 & Prim & B+R & MACHO    \\
 OGLE-SMC-ECL-5233 & 50250.30258 & 0.00997 & Sec  & B+R & MACHO    \\
 OGLE-SMC-ECL-5233 & 50749.41555 & 0.00521 & Prim & B+R & MACHO    \\
 OGLE-SMC-ECL-5233 & 50752.16203 & 0.01977 & Sec  & B+R & MACHO    \\
 OGLE-SMC-ECL-5233 & 51499.37910 & 0.01377 & Prim & B+R & MACHO    \\
 OGLE-SMC-ECL-5233 & 51502.33500 & 0.00936 & Sec  & B+R & MACHO    \\
 OGLE-SMC-ECL-5233 & 56669.66164 & 0.00089 & Prim &  R  & DK154    \\
 OGLE-SMC-ECL-5233 & 56676.65528 & 0.00095 & Sec  &  R  & DK154    \\
  \hline
 OGLE-SMC-ECL-5422 & 48701.27481 & 0.00746 & Prim & B+R & MACHO1+2 \\
 OGLE-SMC-ECL-5422 & 48702.82150 & 0.02491 & Sec  & B+R & MACHO1+2 \\
 OGLE-SMC-ECL-5422 & 49549.49972 & 0.01280 & Prim & B+R & MACHO1+2 \\
 OGLE-SMC-ECL-5422 & 49551.05868 & 0.00682 & Sec  & B+R & MACHO1+2 \\
 OGLE-SMC-ECL-5422 & 49850.48551 & 0.00818 & Prim & B+R & MACHO1+2 \\
 OGLE-SMC-ECL-5422 & 49852.05883 & 0.00491 & Sec  & B+R & MACHO1+2 \\
 OGLE-SMC-ECL-5422 & 50200.10518 & 0.00840 & Prim & B+R & MACHO1+2 \\
 OGLE-SMC-ECL-5422 & 50201.70786 & 0.01002 & Sec  & B+R & MACHO1+2 \\
 OGLE-SMC-ECL-5422 & 50549.72721 & 0.00918 & Prim & B+R & MACHO1+2 \\
 OGLE-SMC-ECL-5422 & 50551.35470 & 0.00736 & Sec  & B+R & MACHO1+2 \\
 OGLE-SMC-ECL-5422 & 50850.70503 & 0.00509 & Prim & B+R & MACHO1+2 \\
 OGLE-SMC-ECL-5422 & 50852.35616 & 0.00474 & Sec  & B+R & MACHO1+2 \\
 OGLE-SMC-ECL-5422 & 51501.30424 & 0.01298 & Prim & B+R & MACHO1+2 \\
 OGLE-SMC-ECL-5422 & 51502.98810 & 0.00716 & Sec  & B+R & MACHO1+2 \\
 OGLE-SMC-ECL-5422 & 50549.70892 & 0.00210 & Prim &  I  & OGLE II  \\
 OGLE-SMC-ECL-5422 & 50551.35676 & 0.00230 & Sec  &  I  & OGLE II  \\
 OGLE-SMC-ECL-5422 & 51099.98977 & 0.00336 & Prim &  I  & OGLE II  \\
 OGLE-SMC-ECL-5422 & 51101.66027 & 0.00092 & Sec  &  I  & OGLE II  \\
 OGLE-SMC-ECL-5422 & 51698.90939 & 0.00154 & Prim &  I  & OGLE II  \\
 OGLE-SMC-ECL-5422 & 51700.61184 & 0.00215 & Sec  &  I  & OGLE II  \\
 OGLE-SMC-ECL-5422 & 52200.54217 & 0.00385 & Prim &  I  & OGLE III \\
 OGLE-SMC-ECL-5422 & 52202.27954 & 0.00557 & Sec  &  I  & OGLE III \\
 OGLE-SMC-ECL-5422 & 52574.48865 & 0.00266 & Prim &  I  & OGLE III \\
 OGLE-SMC-ECL-5422 & 52576.25467 & 0.00306 & Sec  &  I  & OGLE III \\
 OGLE-SMC-ECL-5422 & 52924.11355 & 0.00205 & Prim &  I  & OGLE III \\
 OGLE-SMC-ECL-5422 & 52925.89530 & 0.00320 & Sec  &  I  & OGLE III \\
 OGLE-SMC-ECL-5422 & 53301.10369 & 0.00275 & Prim &  I  & OGLE III \\
 OGLE-SMC-ECL-5422 & 53302.89816 & 0.00329 & Sec  &  I  & OGLE III \\
 OGLE-SMC-ECL-5422 & 53650.71277 & 0.00982 & Prim &  I  & OGLE III \\
 OGLE-SMC-ECL-5422 & 53652.53418 & 0.00438 & Sec  &  I  & OGLE III \\
 OGLE-SMC-ECL-5422 & 54000.35282 & 0.00120 & Prim &  I  & OGLE III \\
 OGLE-SMC-ECL-5422 & 54002.17710 & 0.00006 & Sec  &  I  & OGLE III \\
 OGLE-SMC-ECL-5422 & 54349.98870 & 0.00116 & Prim &  I  & OGLE III \\
 OGLE-SMC-ECL-5422 & 54351.81853 & 0.00001 & Sec  &  I  & OGLE III \\
 OGLE-SMC-ECL-5422 & 54799.94233 & 0.00889 & Prim &  I  & OGLE III \\
 OGLE-SMC-ECL-5422 & 54801.77785 & 0.00402 & Sec  &  I  & OGLE III \\
  \hline
 OGLE-SMC-ECL-5434 & 48798.94606 & 0.01483 & Prim & B+R & MACHO    \\
 OGLE-SMC-ECL-5434 & 48800.47699 & 0.00603 & Sec  & B+R & MACHO    \\
 OGLE-SMC-ECL-5434 & 49800.72760 & 0.00678 & Prim & B+R & MACHO    \\
 OGLE-SMC-ECL-5434 & 49802.25181 & 0.01099 & Sec  & B+R & MACHO    \\
 OGLE-SMC-ECL-5434 & 50199.12527 & 0.01339 & Prim & B+R & MACHO    \\
 OGLE-SMC-ECL-5434 & 50200.65017 & 0.00750 & Sec  & B+R & MACHO    \\
 OGLE-SMC-ECL-5434 & 50551.33912 & 0.00586 & Prim & B+R & MACHO    \\
 OGLE-SMC-ECL-5434 & 50552.84222 & 0.01509 & Sec  & B+R & MACHO    \\
 OGLE-SMC-ECL-5434 & 50848.71070 & 0.00518 & Prim & B+R & MACHO    \\
 OGLE-SMC-ECL-5434 & 50850.18788 & 0.01259 & Sec  & B+R & MACHO    \\
 OGLE-SMC-ECL-5434 & 51501.15782 & 0.00807 & Prim & B+R & MACHO    \\
 OGLE-SMC-ECL-5434 & 51502.62719 & 0.02513 & Sec  & B+R & MACHO    \\
 OGLE-SMC-ECL-5434 & 50551.33644 & 0.00503 & Prim &  I  & OGLE II  \\
 OGLE-SMC-ECL-5434 & 50552.82479 & 0.00705 & Sec  &  I  & OGLE II  \\
 OGLE-SMC-ECL-5434 & 51099.85925 & 0.02141 & Prim &  I  & OGLE II  \\
 OGLE-SMC-ECL-5434 & 51101.33185 & 0.01155 & Sec  &  I  & OGLE II  \\
 OGLE-SMC-ECL-5434 & 51700.34958 & 0.00445 & Prim &  I  & OGLE II  \\
 OGLE-SMC-ECL-5434 & 51701.82145 & 0.00903 & Sec  &  I  & OGLE II  \\
 OGLE-SMC-ECL-5434 & 52199.81176 & 0.00436 & Prim &  I  & OGLE III \\
 OGLE-SMC-ECL-5434 & 52201.23825 & 0.00482 & Sec  &  I  & OGLE III \\
 OGLE-SMC-ECL-5434 & 52575.12842 & 0.00442 & Prim &  I  & OGLE III \\
 OGLE-SMC-ECL-5434 & 52576.53465 & 0.00466 & Sec  &  I  & OGLE III \\
 OGLE-SMC-ECL-5434 & 52924.44326 & 0.00311 & Prim &  I  & OGLE III \\
 OGLE-SMC-ECL-5434 & 52925.84853 & 0.00567 & Sec  &  I  & OGLE III \\
 OGLE-SMC-ECL-5434 & 53299.75304 & 0.00310 & Prim &  I  & OGLE III \\
 OGLE-SMC-ECL-5434 & 53301.15126 & 0.00369 & Sec  &  I  & OGLE III \\
 OGLE-SMC-ECL-5434 & 53649.08118 & 0.00282 & Prim &  I  & OGLE III \\
 OGLE-SMC-ECL-5434 & 53650.46107 & 0.00647 & Sec  &  I  & OGLE III \\
 OGLE-SMC-ECL-5434 & 54001.28546 & 0.00260 & Prim &  I  & OGLE III \\
 OGLE-SMC-ECL-5434 & 54002.65242 & 0.00208 & Sec  &  I  & OGLE III \\
 OGLE-SMC-ECL-5434 & 54350.61859 & 0.00245 & Prim &  I  & OGLE III \\
 OGLE-SMC-ECL-5434 & 54351.97724 & 0.00573 & Sec  &  I  & OGLE III \\
 OGLE-SMC-ECL-5434 & 54800.98529 & 0.00261 & Prim &  I  & OGLE III \\
 OGLE-SMC-ECL-5434 & 54802.33139 & 0.00413 & Sec  &  I  & OGLE III \\
 \hline
\noalign{\smallskip}\hline
\end{tabular}
\end{minipage}
\end{table}

\end{appendix}
\end{document}